%% file: LCWS15.tex
\documentclass[a4paper]{article}
\usepackage{graphicx}
\usepackage{cite}

\def\gevc{$\rm GeV/c$}
\def\gevc2{$\rm GeV/c^2$}

\def\mevc{$\rm MeV/c$}
\def\mevc2{$\rm MeV/c^2$}

\def\Vbias{$V_{bias}$}

\newcommand\pubnumber{}
\newcommand\pubdate{\today}

\def\Title#1{\begin{center} {\Large #1 } \end{center}}
\def\Author#1{\begin{center}{ \sc #1} \end{center}}
\def\Address#1{\begin{center}{ \it #1} \end{center}}

\newcommand\pubblock{\rightline{\begin{tabular}{l} \pubnumber\\
         \pubdate  \end{tabular}}}

\input econfmacros.tex

\begin{document}
\begin{titlepage}
\pubblock

\vfill
\Title{SiPM Gain Stabilization Studies for Adaptive Power Supply}


\vfill
\Author{{\bf Gerald Eigen}, Are Tr\ae et, Justas Zalieckas} 
\Address{Department of Physics and Technology\\
University of Bergen, N-5007 Bergen, Norway 
}
\Author{Jaroslav Cvach, Jiri Kvasnicka, Ivo Polak} 
\Address{Institute of Physics of the ASCR \\
Prague, Czech Republic}

\Author{Talk presented at the International Workshop on Future Linear Colliders (LCWS15), Whistler, Canada, 2-6 November 2015.}

\vfill
\pagenumbering{arabic}

\begin{abstract}
We present herein gain stabilization studies of SiPMs using a climate chamber at CERN. We present results for four detectors not tested before, three from Hamamatsu and one from KETEK. Two of the Hamamatsu SiPMs are novel sensors with trenches that reduce cross talk. We use an improved readout system with a digital oscilloscope controlled with a dedicated LabView program. We improved and automized the analysis to deal with large datasets. We have measured the gain-versus-bias-voltage dependence at fixed temperature and gain-versus-temperature dependence at fixed bias voltage to determine the bias voltage dependence on temperature $V(T)$ for stable gain. We show that the gain remains stable to better than $\pm 0.5\%$ in the $20^\circ \rm C - 30^\circ C$ temperature range if the bias voltage is properly adjusted with temperature.

\end{abstract}

\end{titlepage}

\section{Introduction}

The gain of silicon photomultipliers (SiPMs)~\cite{Bondarenko, Buzhan, Buzhan2} increases with bias voltage (\Vbias) and decreases with temperature (T). For stable operation, the gain needs to be kept constant, especially in large detector systems such as an analog hadron calorimeter operating of the order of $10^6$ SiPMs\cite{ilc}. The method for keeping the gain constant consists of adjusting \Vbias\ when T changes. This, however, requires knowledge of $dV/dT$, which can be extracted from measurements of the gain-versus-bias-voltage (dG/dV)  dependence and that of the gain versus temperature (dG/dT). Within the AIDA framework, we measured $dG/dV$ and $dG/dT$ for 17 SiPMs from three manufacturers (Hamamatsu, KETEK, CPTA) in a climate chamber at CERN. 
The goal was to demonstrate gain stability within $\pm 0.5\%$ for individual SiPMs within the $20^\circ \rm C - 30^\circ C$ temperature range. Thus, we selected four of the 17 SiPMs and demonstrated gain stabilization to $< \pm 0.5\%$ for these~\cite{cvach}. In the AIDA2020 framework, we want to show that with one bias voltage adjustment the gain of several similar SiPMs can be kept stable. The goal is to implement the temperature-dependent bias voltage adjustment into the power distribution system of the analo                                                                                                                                                   g hadron calorimeter~\cite{ahcal}. 

In August 2015, we conducted our first set of measurements within the AIDA2020 framework using a new, improved readout system. We performed $dG/dV$ and $dG/dT$ measurements for several SiPMs we previously tested as well as for a couple of novel Hamamatsu SiPMs with trenches. For gain stabilization, we focused first  on SiPMs we did not test previously using the same \Vbias\ regulator board as before. We also wanted to crosscheck the CPTA sensors since their photoelectron spectra contained much more background. One complication with the CPTA sensors is that they are mounted to a scintillator tile and we had to inject light through the tile and not directly to the SiPM. We built a setup to test two SiPMs simultaneously. Unfortunately, one channel did not work properly  due to an unreliable performance of one preamplifier.

\section{Gain Calibration Test Setup}

\begin{figure}[h]
\centering
\vskip -0.2cm
\includegraphics[width=80mm]{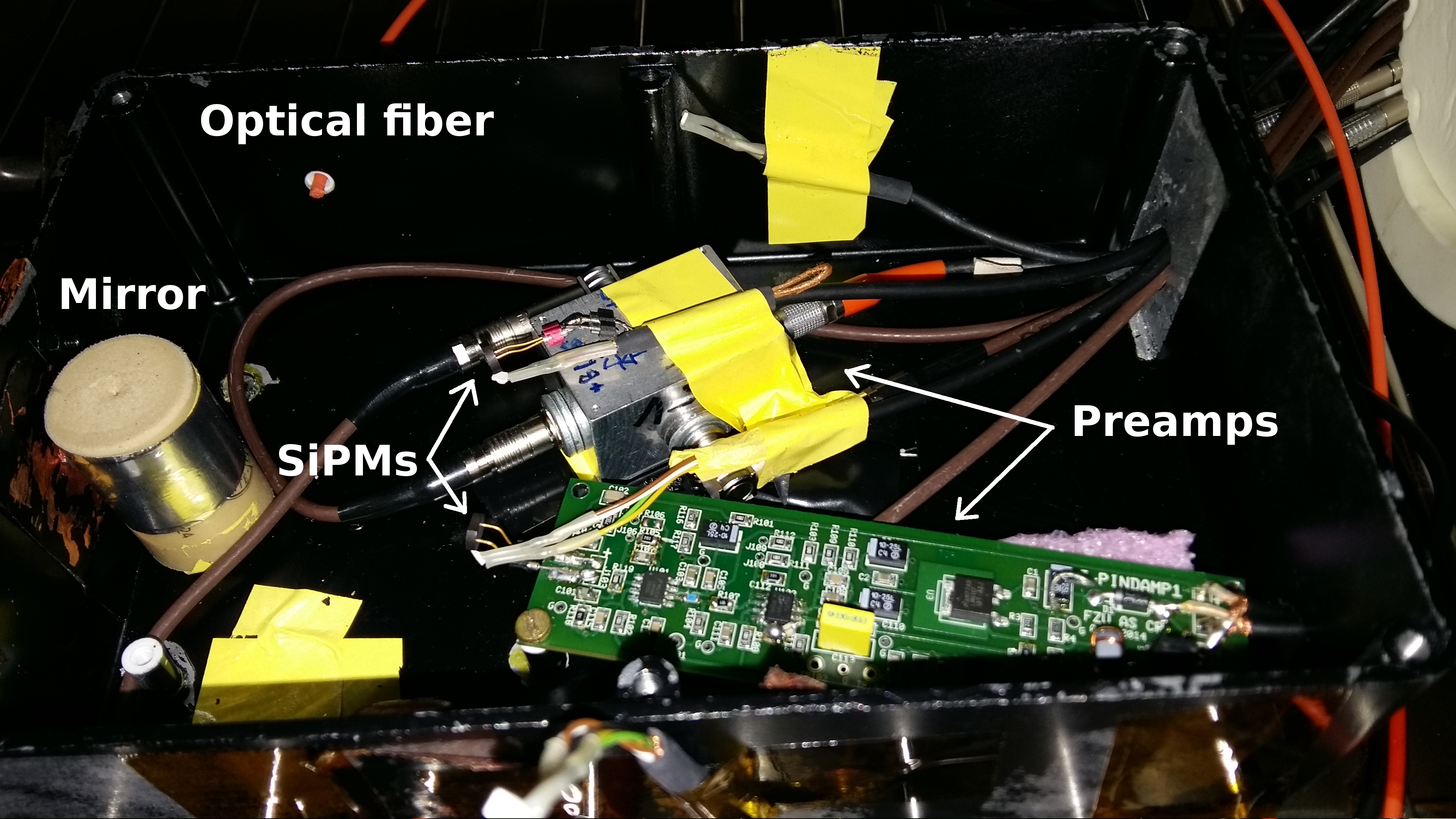}
\includegraphics[width=70mm]{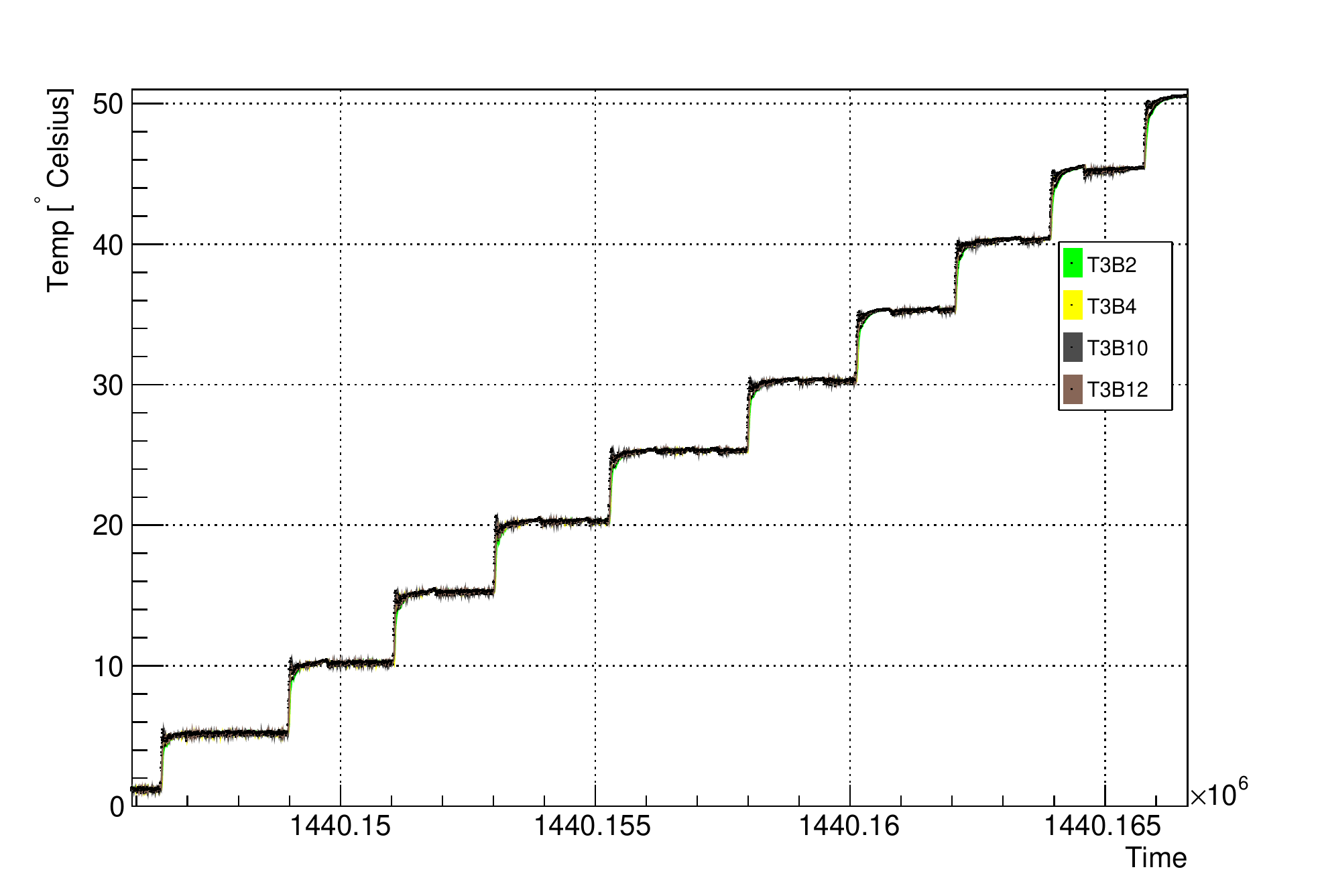}
\caption{(Left) Setup of the gain stabilization measurements inside the black box showing two SiPMs, two preamplifiers, four temperature sensors (fixed with yellow tape), the optical fiber and the mirror. (Right) Typical temperature profile during the gain stabilization measurements in seconds.}
\label{fig:setup}
\end{figure}

Figure~\ref{fig:setup} (left) shows our measurement setup in the climate chamber Spiral3 from Climats at CERN. Two SiPMs, mounted inside a black box in the climate chamber, were connected to two different preamplifiers, one voltage-operational two-stage preamplifier (8/25 ns) and one current-operational one-stage preamplifier (2/7 ns). Since only the voltage-operational preamplifier performed properly at all temperatures between $1^\circ$C and $50^\circ$C, we could use only that one for gain stability studies. One PT1000 temperature sensor was placed close to each SiPM, a third PT1000 sensor was attached to the wall of the black box, a fourth was fixed inside the black box and a fifth was used to monitor the temperature inside the climate chamber.
The climate chamber shows a temperature offset of $+0.5^\circ$C that remains unchanged over the entire $1^\circ$C to $50^\circ$C temperature range.
However, for scans with increasing temperatures the offset is positive while for those with decreasing temperature it is negative. In the gain stability tests, we typically went from low to high temperatures. 
After reaching equilibrium, the temperature remains stable within $\sim \pm 0.2^\circ$C. The SiPMs are illuminated simultaneously with blue light transported from an LED  via an optical fiber and a mirror inside the black box. The LED was trigged by a light pulser placed outside the climate chamber to reduce  noise pickup. The light pulser signal is based on a sinusoidal pulse above a selected threshold producing 3.6 ns wide signals. The repetition rate and light intensity are adjustable. We run at 10 kHz. We record SiPM waveforms directly with a digital oscilloscope from LeCroy (model 6104, 12 bit ADC, 2.5 GS/s, 4 channels). Offline, we convert the waveforms into photoelectron spectra after subtracting a DC offset and integrating 50000 waveforms over a 74 ns wide time window. The photoelectron spectra typically show well-separated individual photoelectron $(pe)$ peaks. We wrote a dedicated LabView program for data taking, setting the intensity of the light pulser and controlling the bias voltages of the SiPMs. The low voltages of the preamplifier are set manually and the temperature profiles are recorded by a separate dedicated system built by MPI Munich\cite{munich}. 
We varied the temperature from $5^\circ$C to $45^\circ$C in steps of $5^\circ$C. In the $20^\circ - 30^\circ$C temperature range we reduced the step size to $2^\circ$C.  Figure~\ref{fig:setup} (right) shows a typical temperature profile used in the gain stabilization studies. 
The properties of the SiPMs we tested are listed in Table~\ref{tab:sipm}.

\begin{table}
\centering
\caption{Properties of SiPMs studied.  } \medskip
{\footnotesize
\begin{tabular}{|c|c|c|c|c|c|c|c|c|}
 \hline \hline 
SiPM & Manufacturer & sensitive area $\rm  [mm^2] $& Pitch [$\mu \rm m$]& $\# Pixels$ &$V_{bias} [V] $ & $V_{break} [V] $& Gain [$10^6$] \\
\hline
B2 & Hamamatsu & $1 \times 1$&15 &  4440 &73.99& 71.3 & 0.26    \\
LCT4$\#6$ & Hamamatsu &   $1 \times 1$& 50&400 & 53.81 &50.81 &1.6 \\
LCT4$\#9$ & Hamamatsu &$1 \times 1$ & 50 &400 & 53.98 & 50.98& 1.6  \\
W12 & KETEK &$3 \times 3$ & 20& 12100 & 28 &25 & 0.54 \\
 \#857 &CPTA & $1 \times 1$& 40 &796 & 33.4& 31.9& 0.7 \\
\hline \hline                        
\end{tabular}
}
\label{tab:sipm}
\end{table}

\section{Gain Determination}

\begin{figure}[h]
\centering
\vskip -0.2cm
\includegraphics[width=70mm]{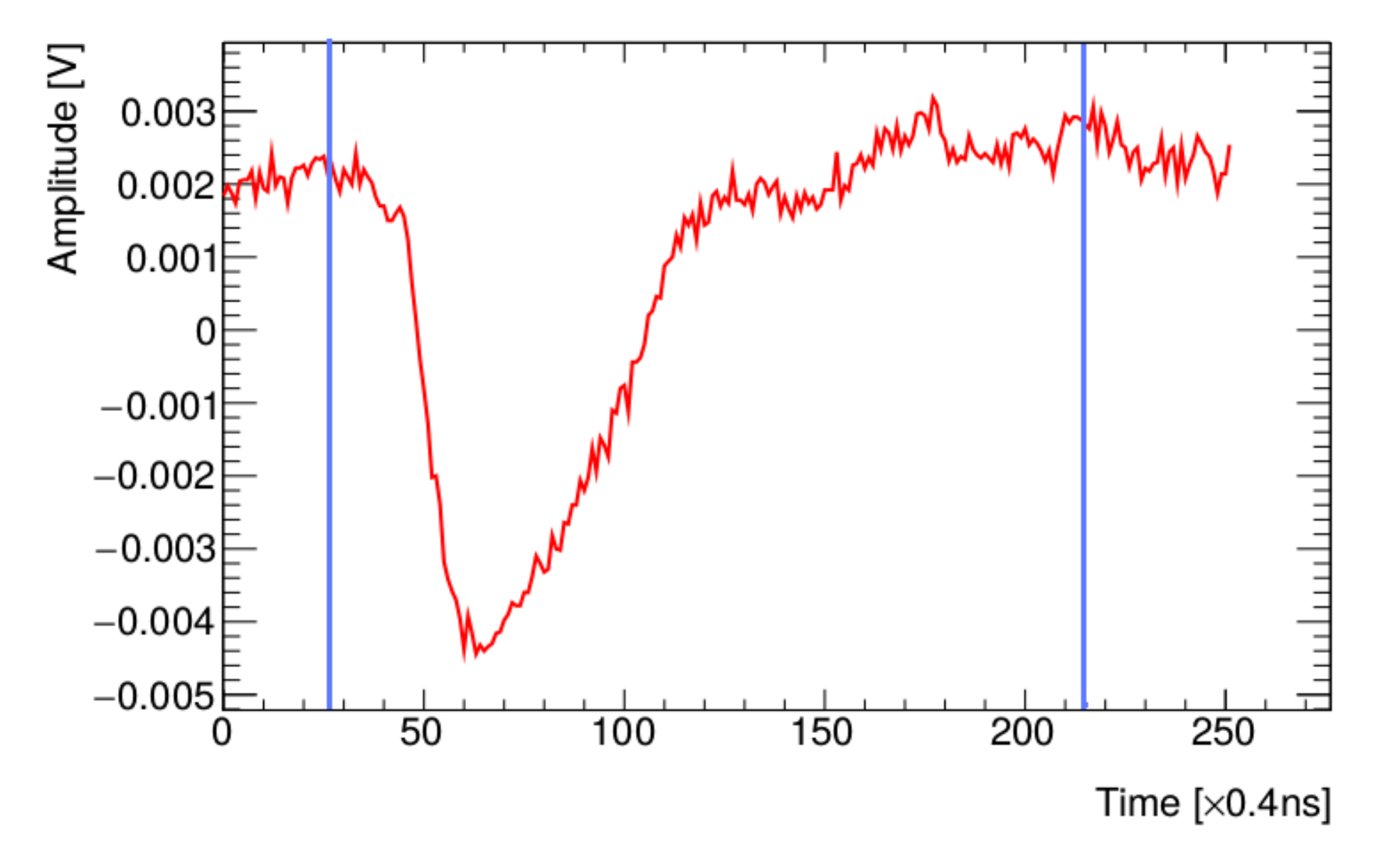}
\includegraphics[width=75mm]{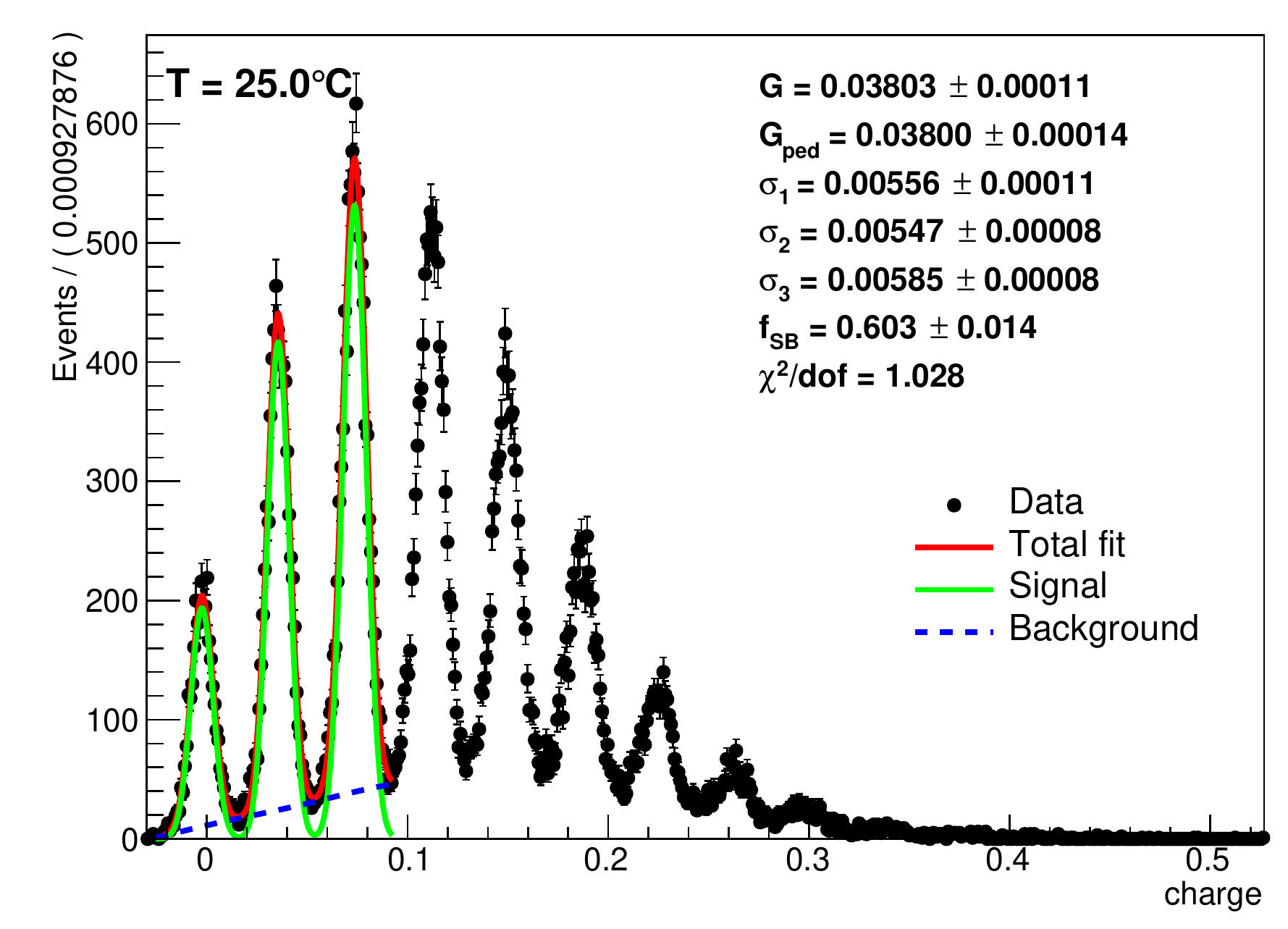}
\caption{A typical waveform (left) and the corresponding photoelectron spectrum (right) for SiPM~B2.}
\label{fig:pe1}
\end{figure}

Figure~\ref{fig:pe1} (left) shows a typical waveform for Hamamatsu MPPC B2 at $25^\circ \rm C$. Figure~\ref{fig:pe1} (right) shows the corresponding photoelectron spectrum. We fit the $pe$ spectrum using the likelihood function:
\begin{equation}
{\cal L}= \prod\limits_{n=1}^{50000} f_s F_{sig}(w^i) +(1-f_s) F_{bkg}(w^i),
\end{equation}
where $f_s$ is the signal fraction. For signal, we model the probability density function (PDF)  with three
Gaussians, one for the pedestal, one for the first $pe$ peak and another for the second $pe$ peak: 
  \begin{equation}
F_{sig} =f_{ped} G_{ped} + f_1 G_1+ (1-f_1 -f_{ped}) G_2.
\end{equation}
The background PDF is parameterized by a sensitive nonlinear iterative peak (SNIP) clipping algorithm that is implemented in the ROOT T-spectrum class. Background is produced by noise,  dark rate, crosstalk  and afterpulsing. The latter originates from pulses triggered by a previous avalanche. Since these pulses are delayed with respect to the original signal, their charge is integrated only partially due to the fixed integration window.
We perform binned fits of the spectra, which have at least two visible photoelectron peaks plus the pedestal. The gain is determined from the distance between the first and second photoelectron peaks. This method is more reliable than using the distance between the pedestal and the first photoelectron peak, which is often smaller than the distance between first and second photoelectron peaks. The statistical error on the gain is obtained from uncertainties of the peak positions of both photoelectron peaks, $\sigma_{gain}=\sqrt{\sigma_1^2 + \sigma_2^2}$. The parameters of three Gaussian functions are not constrained in the fit. At higher bias voltage, the photoelectron spectra show more background while for similar bias voltage the spectra at $5^\circ$C and $45^\circ$C look similar as those at $25^\circ$C.

\section{Measurements for the Hamamatsu SiPM B2}

The SiPM B2 sensor is a $\rm 1~mm \times 1~mm$ SiPM with a pitch of $15~\mu m$. The nominal operation voltage at $25^\circ$C is $V_{bias} = 73.99$ V.

\subsection{Determination of $dG/dV$, $dG/dT$ and $dV/dT$}

\begin{figure}[h]
\centering
\vskip -0.2cm
\includegraphics[width=75mm]{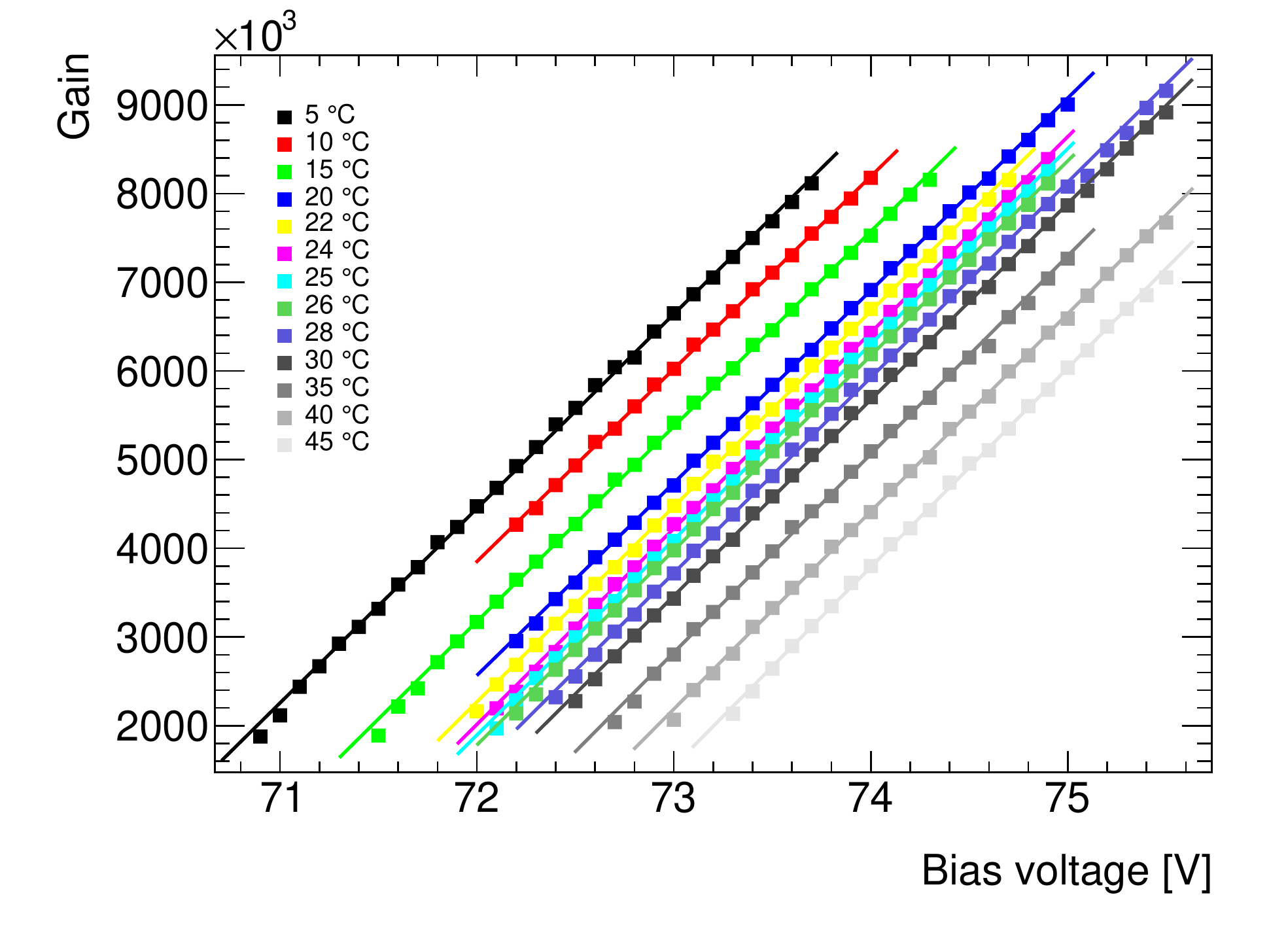}
\includegraphics[width=75mm]{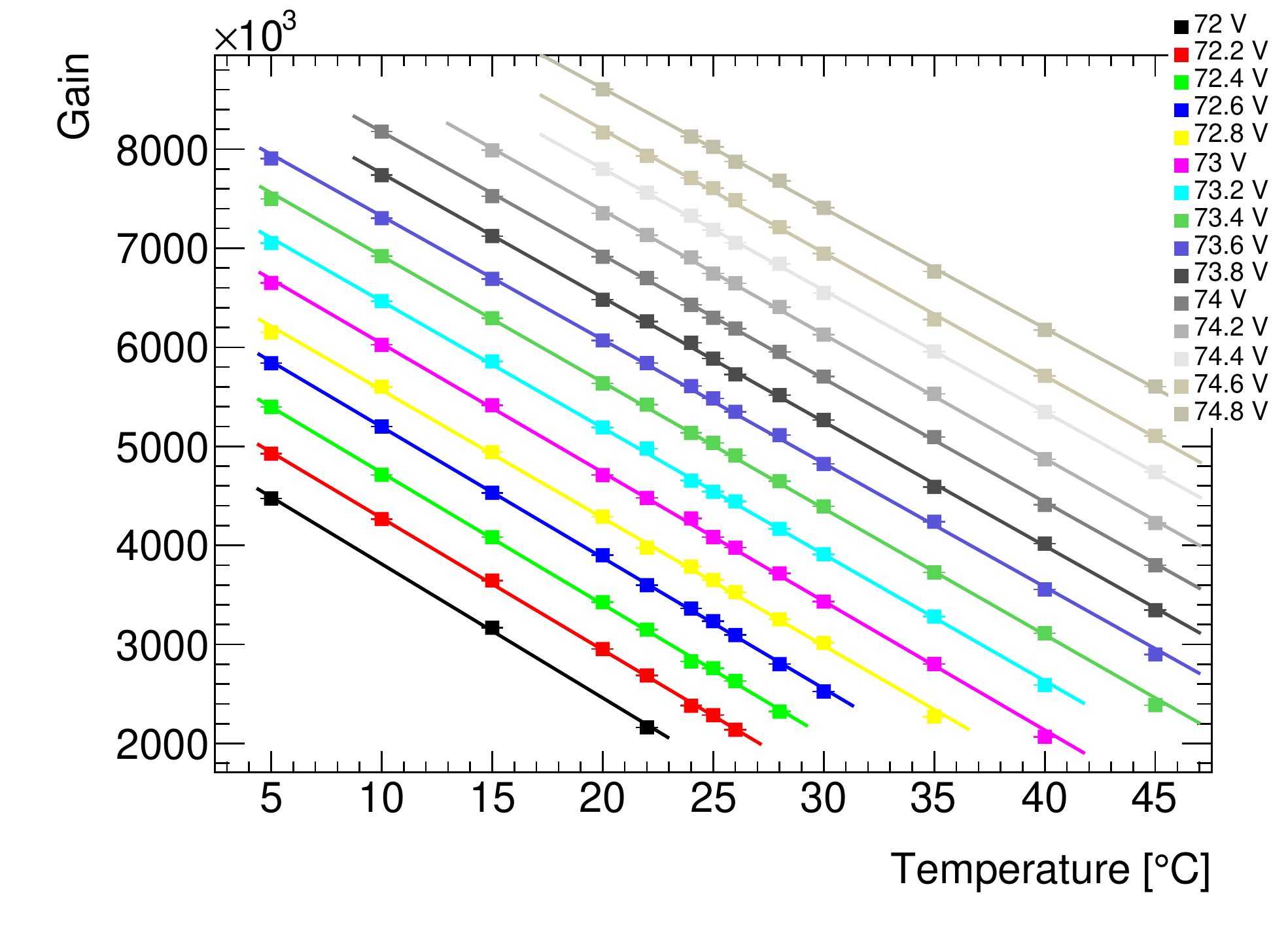}
\caption{Distributions of gain versus bias voltage for fixed temperatures (left) and gain versus temperature for fixed bias voltages (right) for SiPM B2.}
\label{fig:dgdv-b2}
\end{figure}

\begin{figure}[h]
\centering
\vskip -0.2cm
\includegraphics[width=50mm]{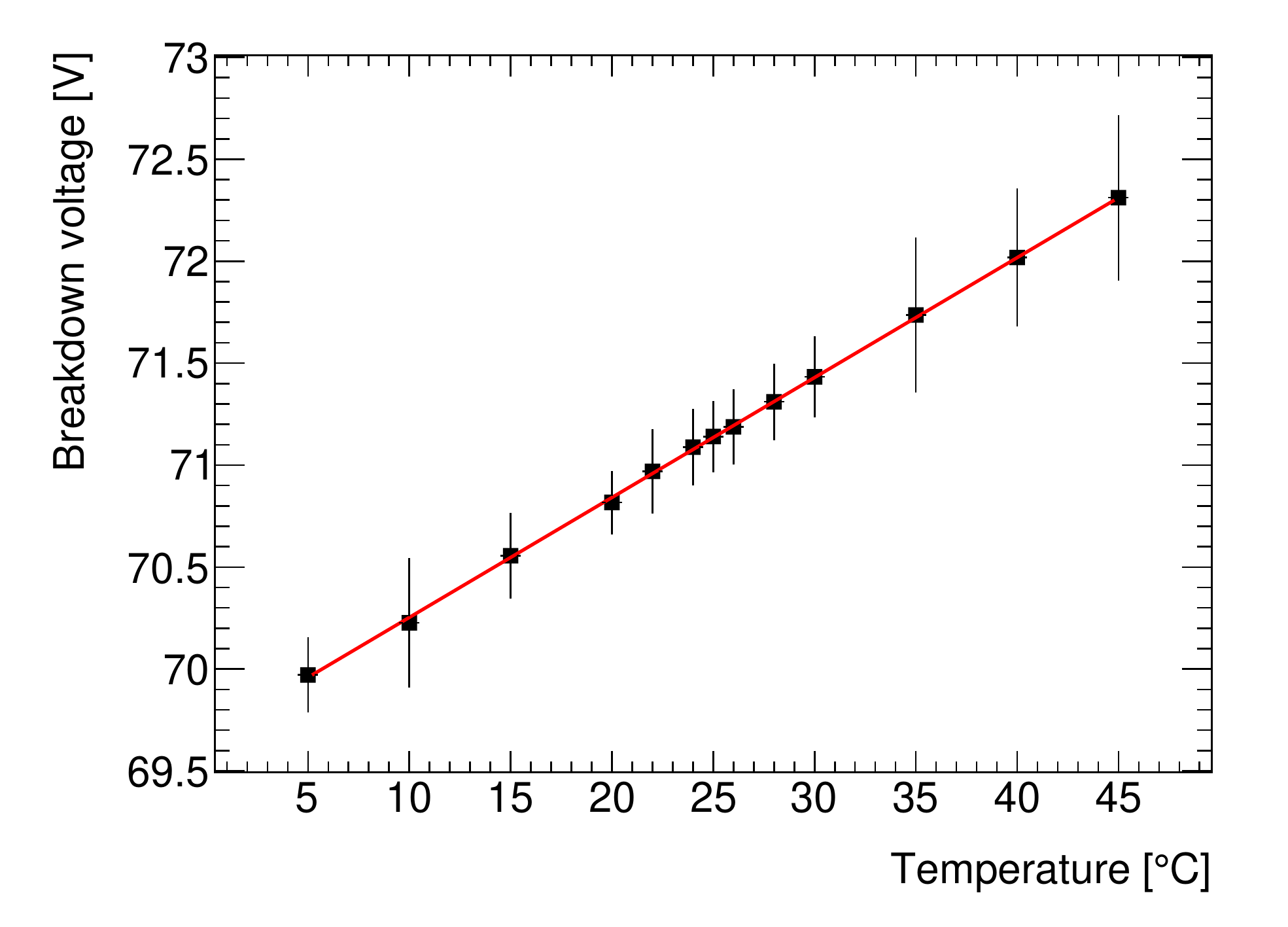}
\includegraphics[width=50mm]{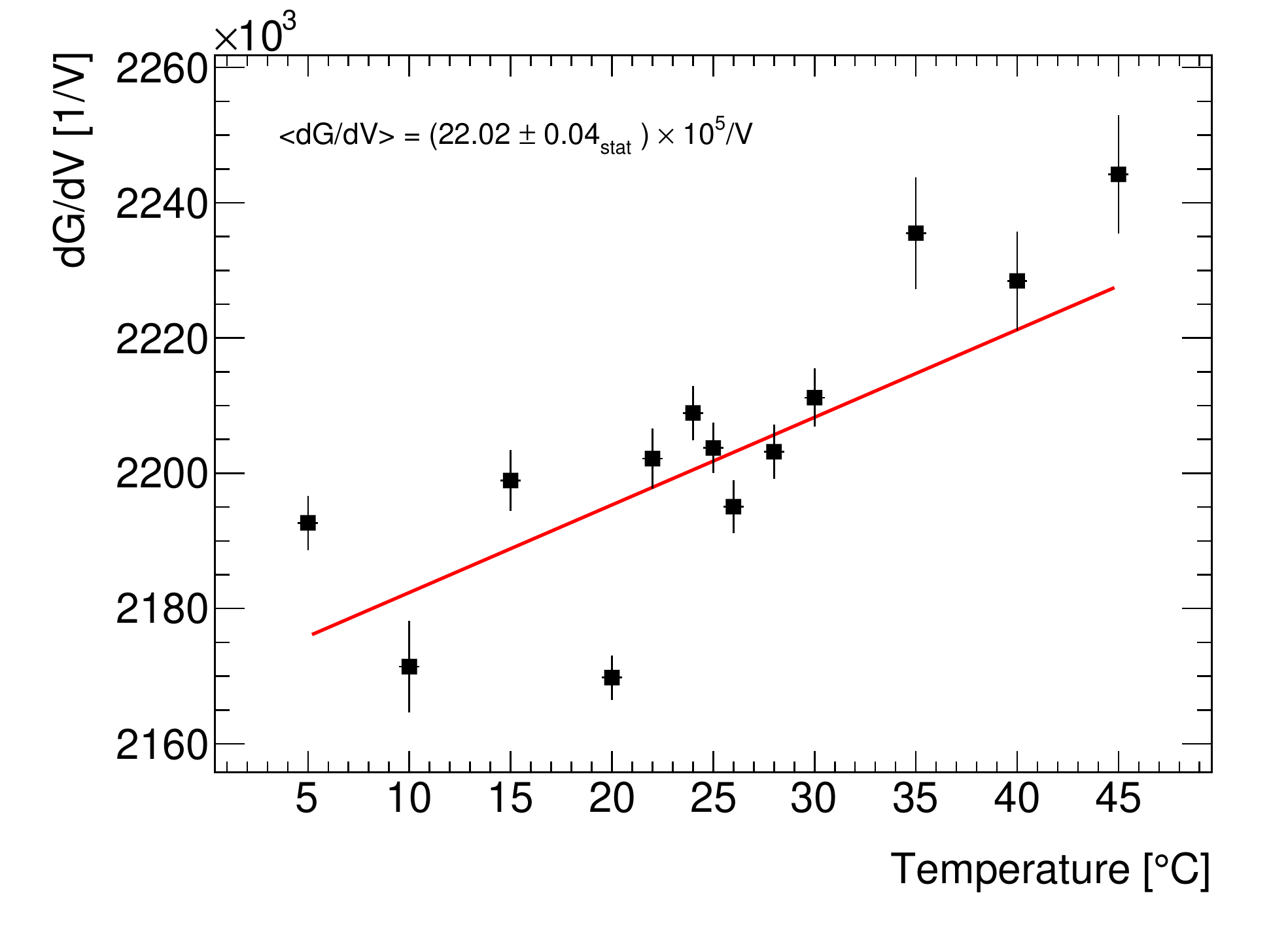}
\includegraphics[width=50mm]{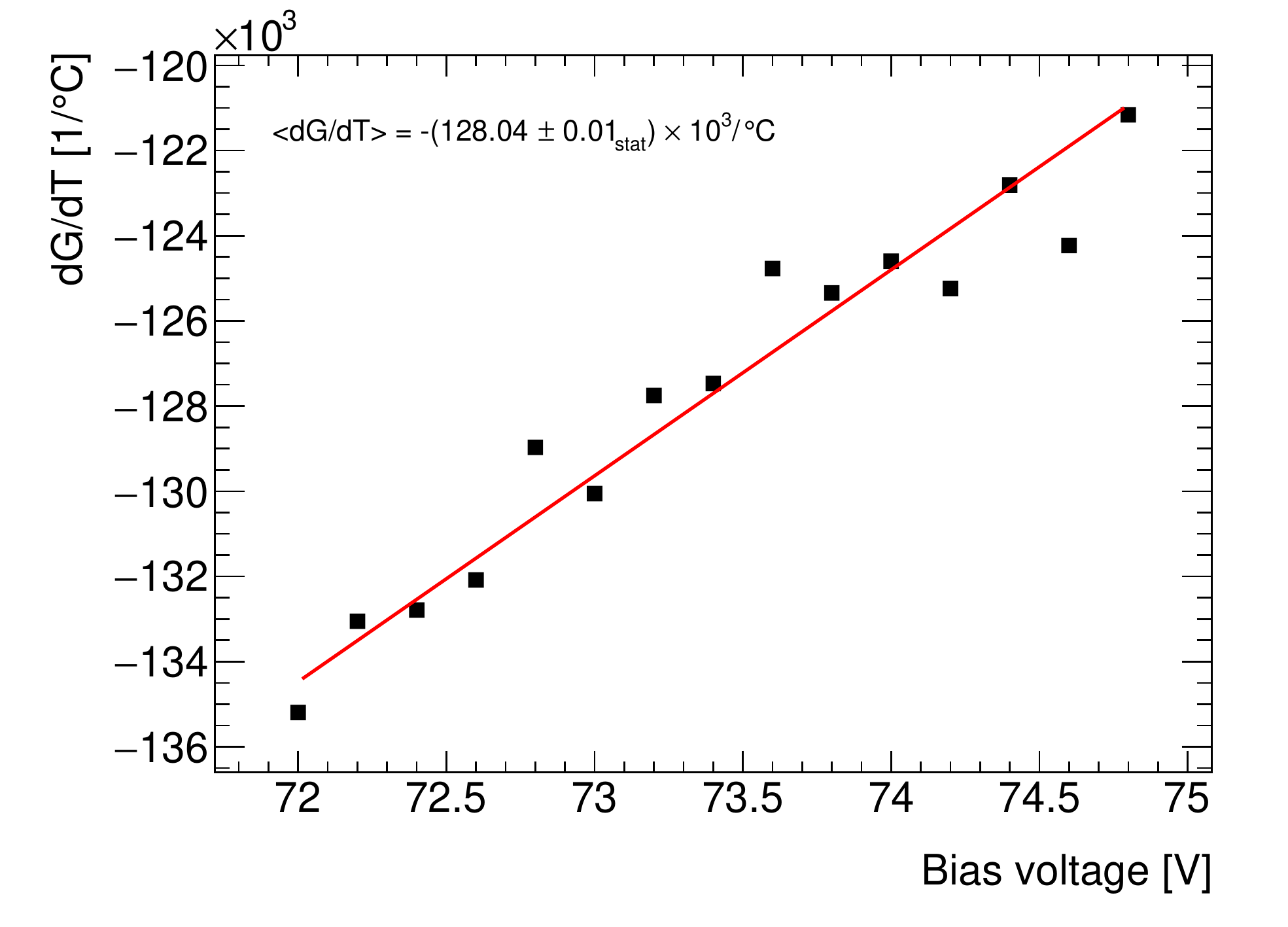}
\caption{Distributions of the break down voltage versus temperature (left), $dG/dV$ versus temperature (middle) and $dG/dT$ versus bias voltage (right) for SiPM B2.}
\label{fig:break-b2}
\end{figure}

\begin{figure}[h]
\centering
\vskip -0.2cm
\includegraphics[width=70mm]{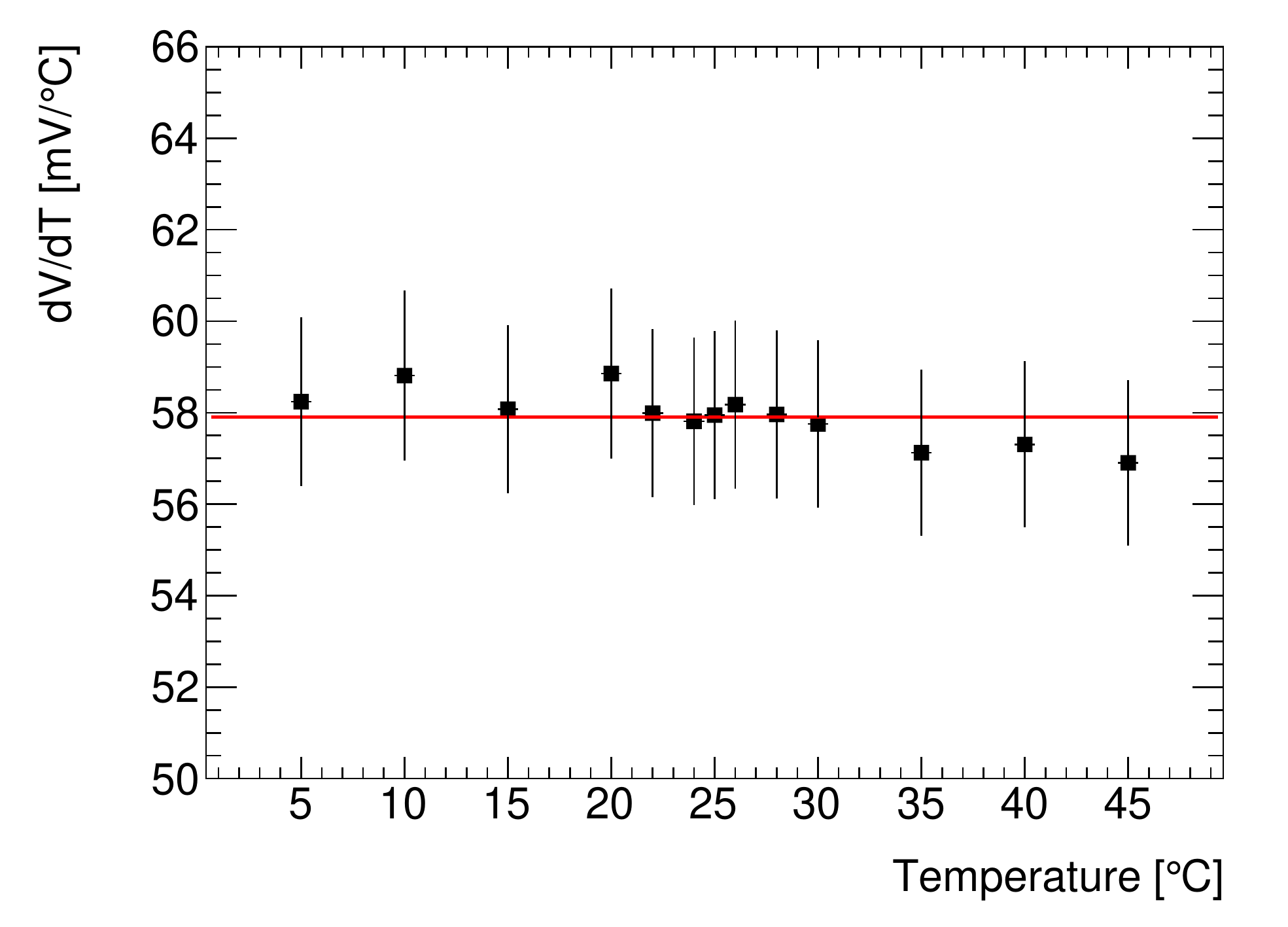}
\includegraphics[width=70mm]{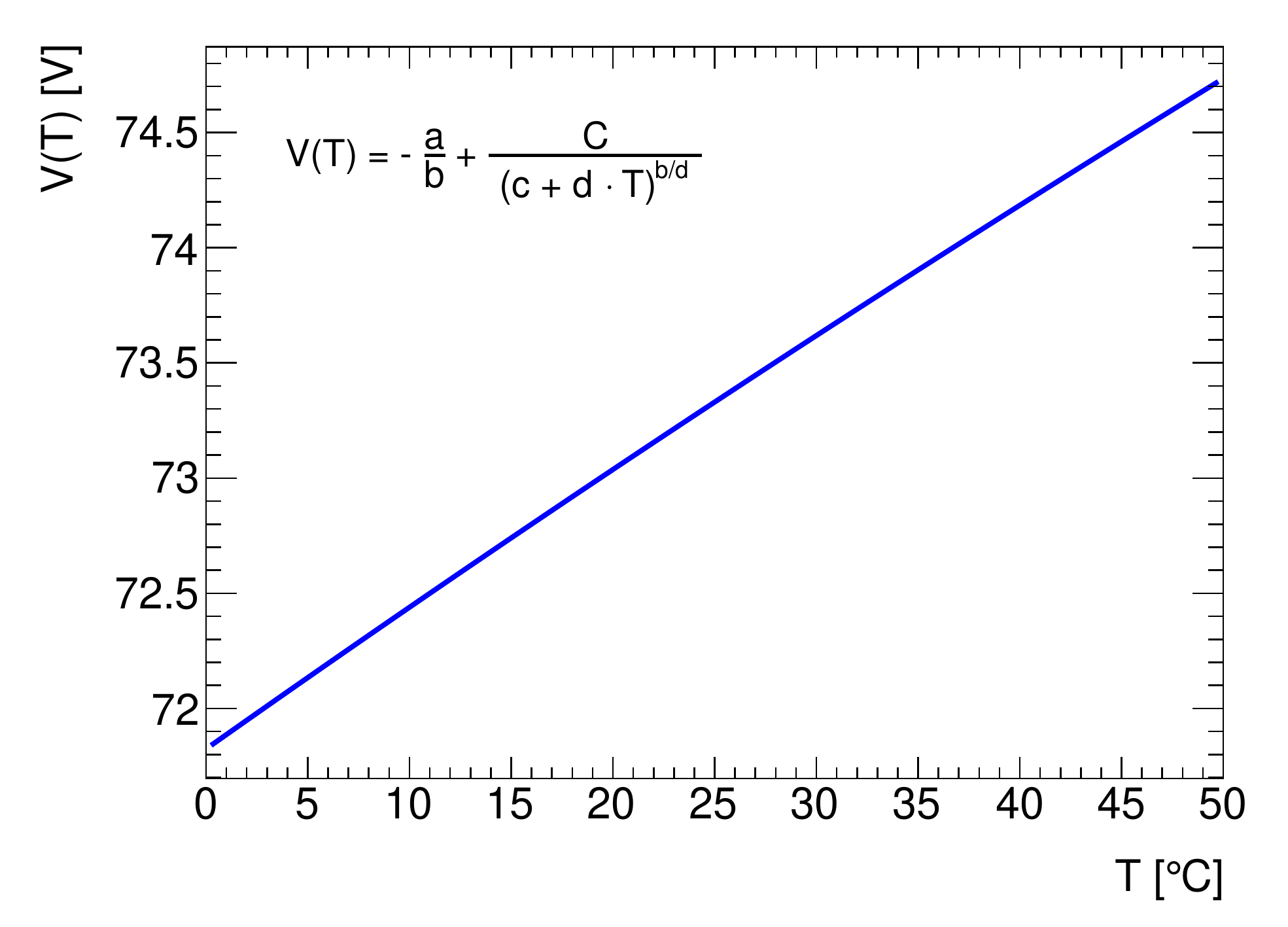}
\caption{Distributions of $\overline{dV/dT}$ versus temperature (left) and $V(T)$ versus temperature (right) for SiPM B2.} 
\label{fig:dvdt-b2}
\end{figure}

First, we measure the gain-versus-bias-voltage dependence at fixed temperatures in 0.1 V steps recording 50000 waveforms at each point. We typically scan over a $\pm1.5$~V region symmetrically around the recommended operational bias voltage at the selected temperature. 
 At each temperature, we fit the measurements with a linear function to extract  the slope $dG/dV$ and the offset. The latter is a linear extrapolation of the break-down voltage $V_{break}$. Figure~\ref{fig:dgdv-b2} (left) shows the results for SiPM~B2 with fits overlaid. All fit curves are nearly parallel. Figure~\ref{fig:break-b2} (left) and (middle) show $V_{break}$ and $dG/dV$ as a function of temperature, respectively.  For SiPM B2, both quantities increase linearly with temperature. 
Note that $dG/dV$ is proportional to the capacitance of the SiPM indicating that the capacitance increases with temperature, which was reported previously~\cite{Dinu}. In the $5^\circ \rm C - 45^\circ$C temperature range, the effect is about 2\%. At the nominal bias voltage, we measure  $dG/dV = (2.202\pm 0.004_{stat})\times 10^6/$V. Next,  we determine the gain-versus-temperature dependence for fixed bias voltage. The gain decreases with temperature. For each value of \Vbias, we fit a linear function to the data to extract $dG/dT$. Figure~\ref{fig:dgdv-b2} (right)  depicts our measurements  for SiPM B2 with fit results overlaid. Figure~\ref{fig:break-b2} (right) shows the resulting $dG/dT$ values as a function of bias voltage. The data reveal a linear dependence on \Vbias. The variation is 11\% in the $5^\circ - 45^\circ$C temperature range. At $25^\circ$C, the fit yields $dG/dT =-(0.12804\pm 0.00001)\times 10^6/^\circ$C. Dividing $dG/dT$ by $dG/dV$ at $25^\circ$C, yields  $dV/dT  = 58.15\pm 0.1\rm  mV/^\circ$C. This value is slightly smaller than the specification of 60 mV/$^\circ$C quoted by Hamamatsu. To estimate the measurement systematic uncertainties, we combine all $dG/dV$\ and $dG/dT$ measurements at a given temperature and determine mean value $\overline{dV/dT} $ and its standard deviation. We fit the resulting  $\overline{dV/dT} $ values and their uncertainties with a uniform distribution to determine the overall mean $\langle dV/dT \rangle$. Figure~\ref{fig:dvdt-b2} (left) shows the resulting $\langle dV/dT \rangle$ distribution. The fit yields $\langle dV/dT \rangle =-57.9\pm 0.5\rm  mV/^\circ$C.  From this, we estimate a gain stability $\frac{\Delta T}{G} \frac{dG}{dV} \sigma(dV/dT) =0.01\%$in the $20^\circ \rm C - 30^\circ$C temperature range.

\subsection{Analytical Determination of \Vbias\ versus Temperature}

The V(T) distribution can be calculated analytically. The gain $G(T,V)$ is a function of temperature and bias voltage. Thus, a gain change is given by:
\begin{equation}
dG(T,V)=\frac{\partial G}{\partial T} dT+ \frac{\partial G}{\partial V} dV
\end{equation}
For stable gain,$dG=0$ yielding
\begin{equation}
dV/dT=-\frac{\partial G(V,T)/\partial T}{\partial G(V,T)/\partial V}.
\end{equation}
Our measurements indicate linear dependences for $dG/dV$ and $dG/dT$. So we can parameterize these distributions by
\begin{eqnarray}
\partial G(V,T)/\partial T& =& a + b \cdot V  \\
\partial G(V,T)/\partial V& = &c + d\cdot T
\end{eqnarray}
where $ a $ and $ c$ are offsets and $b$ and $d$ are slope parameters. All are determined from the fits.
The general solution is a rational function yielding (for $b \neq 0$ and $d\neq 0$) 
\begin{equation}
V=-\frac{a}{b} +\frac{C}{(c +d\cdot T)^{b/d}} \hskip 2cm (C: \rm integration ~constant).
\end{equation}
For SiPM B2, we measure $ a=(-0.48266\pm 0.00002)\times 10^6,  b=4835.9\pm 0.3, c=(2.169\pm 0.0004)\times 10^6$ and $ d=1295\pm 152$. Figure~\ref{fig:dvdt-b2} (right) shows the resulting $V(T)$ function in the $0^\circ \rm C -50^\circ$C temperature range, which is well approximated by a linear function.

\subsection{Gain Stabilization}

\begin{figure}[h]
\centering
\vskip -0.2cm
\includegraphics[width=70mm]{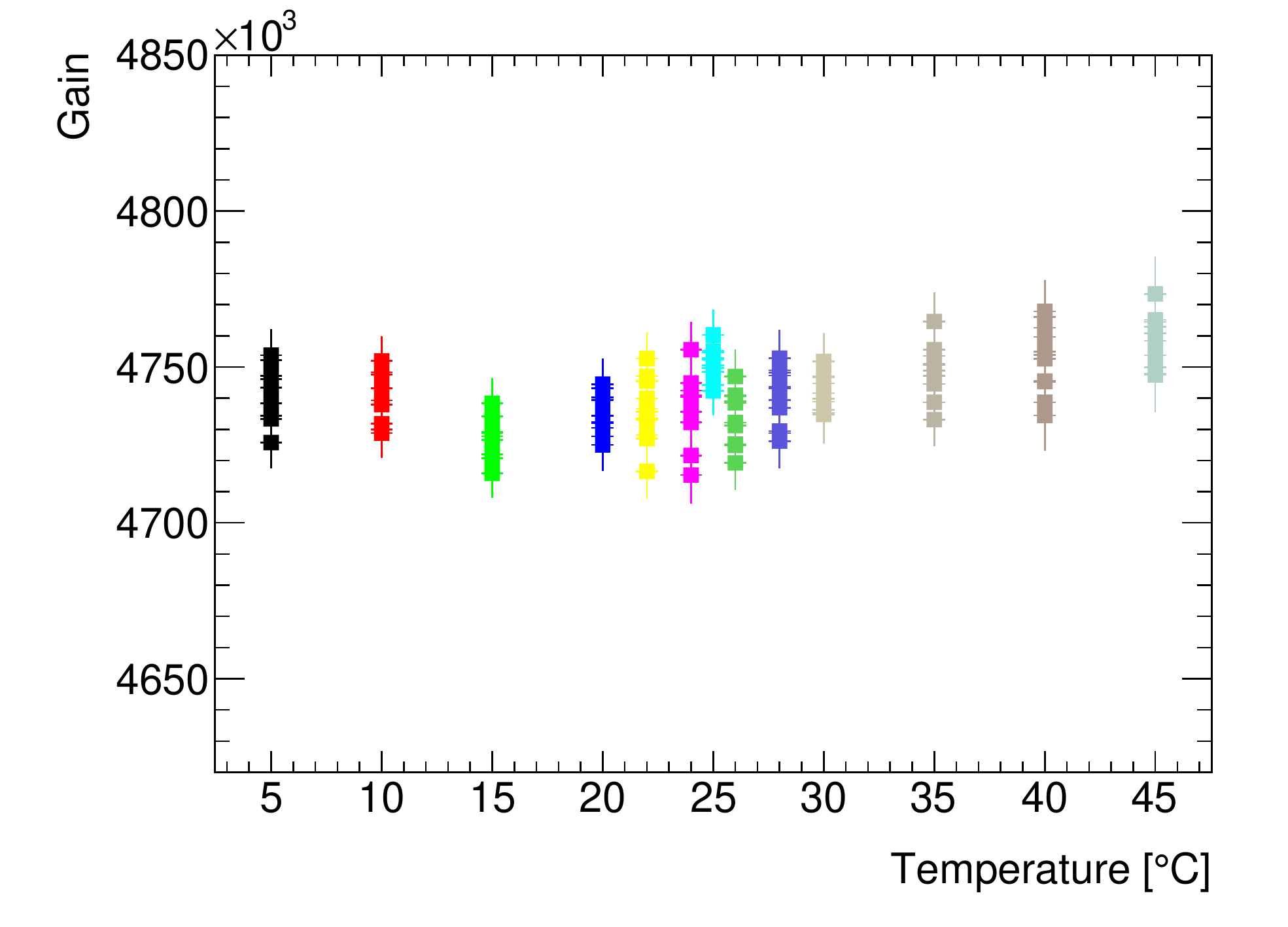}
\includegraphics[width=70mm]{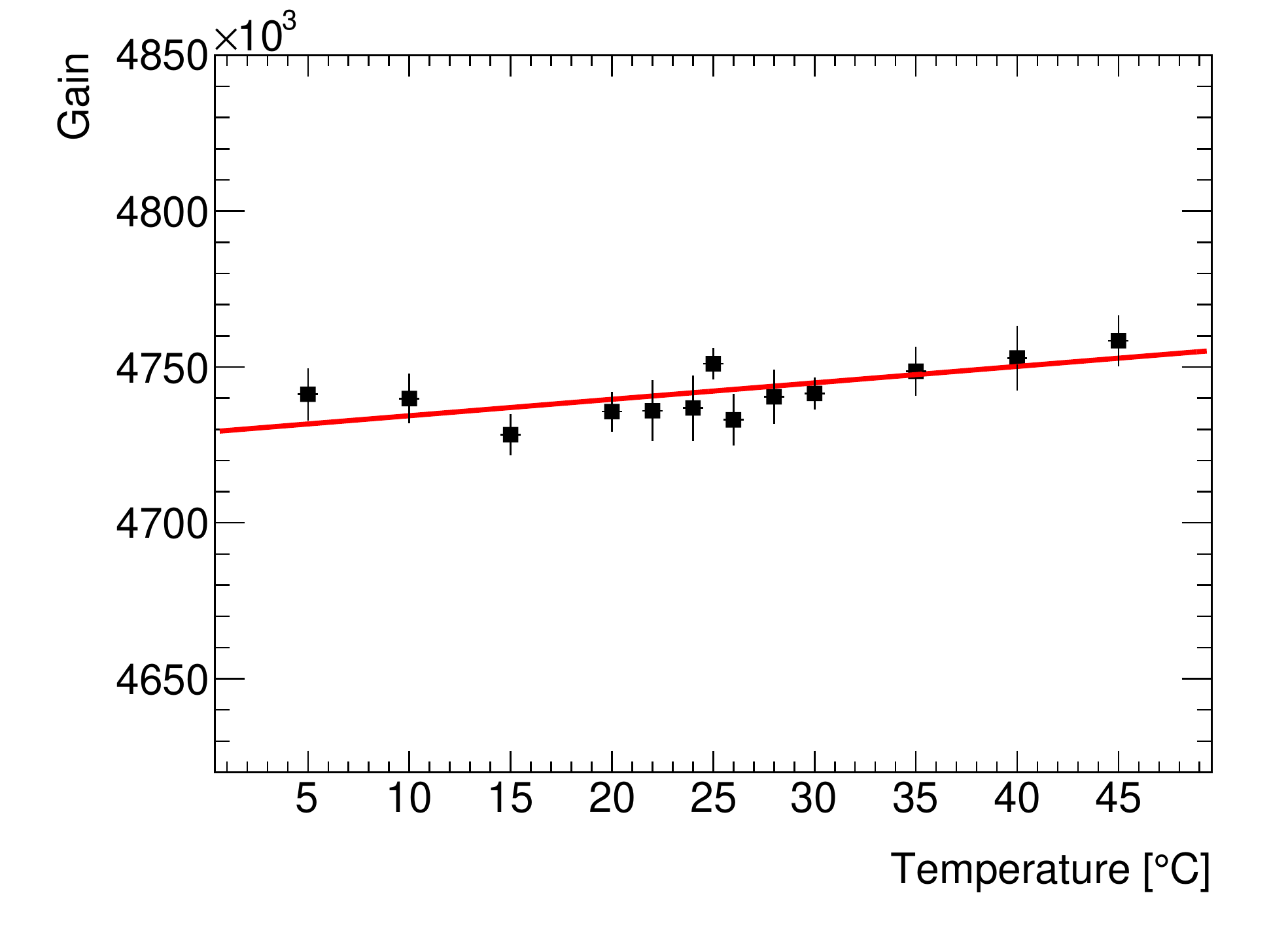}
\caption{Distributions of gain versus temperature for ten measured points at each temperature (left) and for the average of the ten points (right) after stabilization for SiPM B2.} 
\label{fig:stabilization-b2}
\end{figure}

The bias voltage regulator board was built in Prague after thorough tests of a prototype in our previous gain stabilization studies at CERN. Using a compensation of $58~ mV/^\circ $C, the gain stability of  SiPM B2 was tested for 13 temperature points in $5^\circ \rm C - 45^\circ$C temperature range. At each temperature point, ten samples with 50000 waveforms each were recorded. Figure~\ref{fig:stabilization-b2} (left) shows the individual measurements at each temperature point while Fig.~\ref{fig:stabilization-b2} (right) shows the  average over the ten measurements plus its standard deviation. A fit with a linear function yields an offset of $(4.73\pm 0.01)\times 10^6$ and a slope of $527\pm 209$\footnote{Note that the gain is given in arbitrary units since the gain of the preamplifier is unknown.}. In the $5^\circ \rm C - 45^\circ$C temperature range, the gain is rather uniform. The deviation from uniformity is less than $\pm 0.1\%$, which is much smaller than the anticipated gain stability of $\pm0.5\%$ in the $20^\circ  - 30^\circ$C temperature range.

\section{Measurements for the Hamamatsu SiPMs LCT4}

\begin{figure}[h]
\centering
\vskip -0.2cm
\includegraphics[width=70mm]{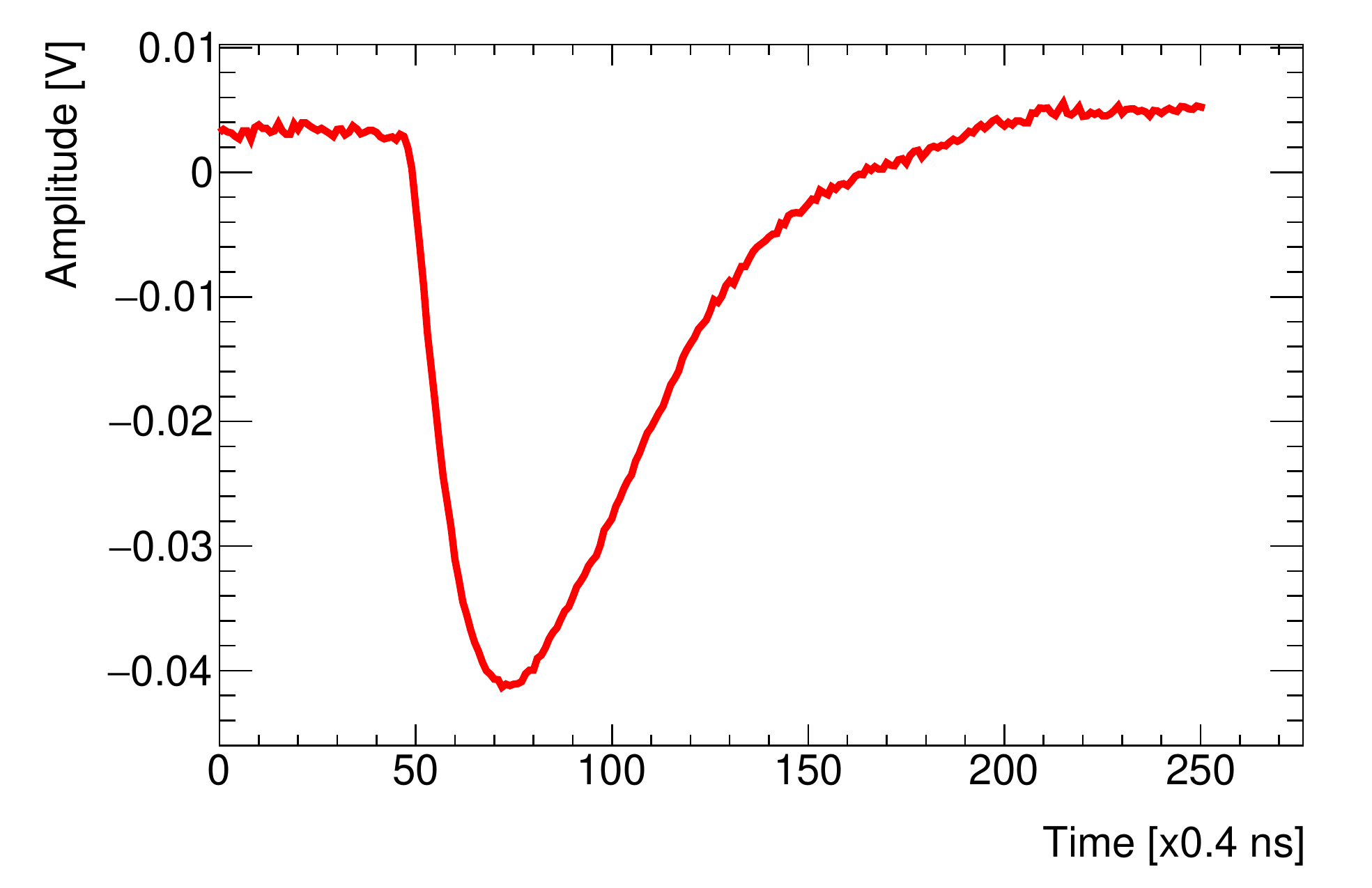}
\includegraphics[width=70mm]{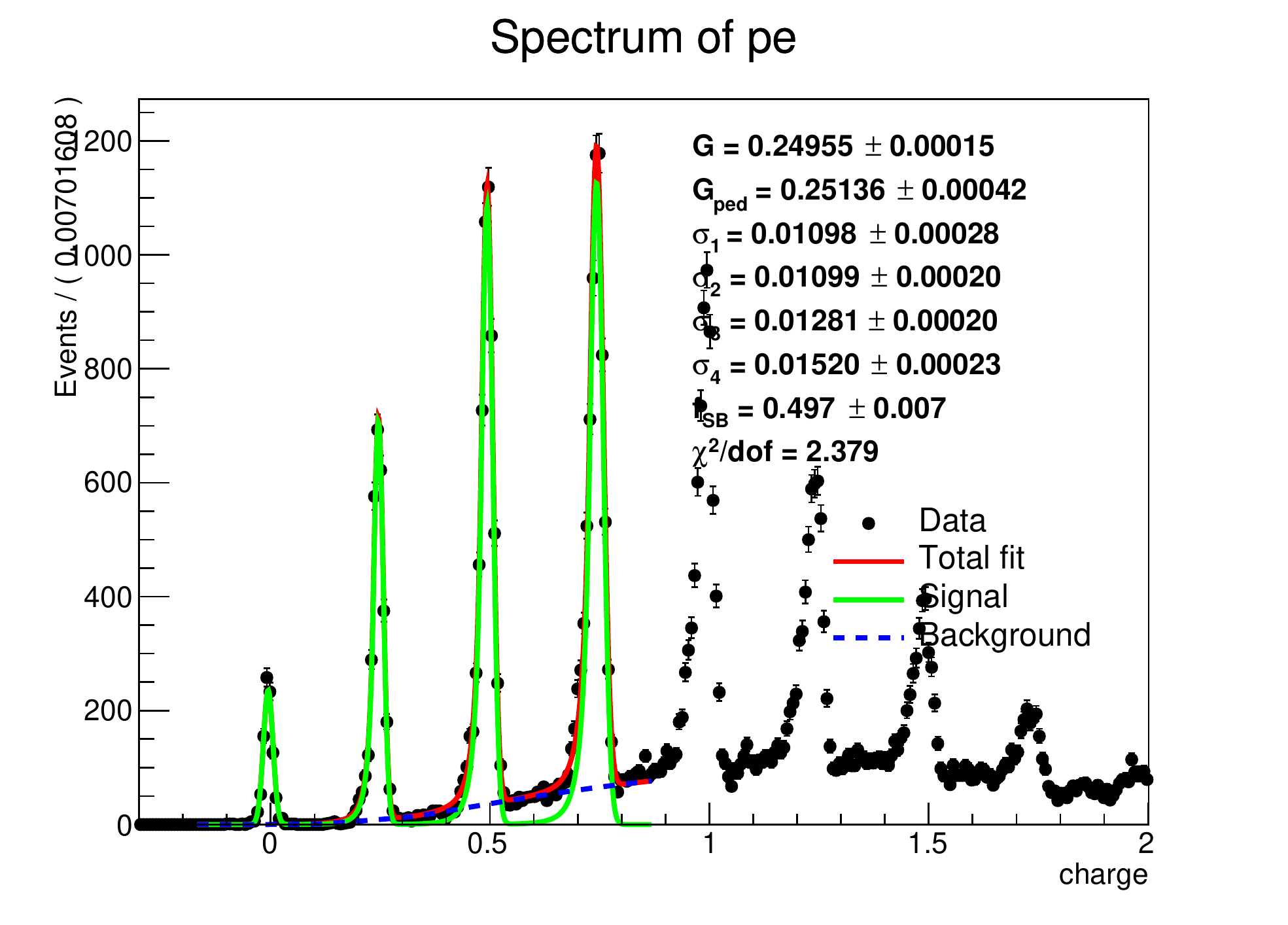}
\caption{A typical waveform (left) and the corresponding photoelectron spectrum (right) for SiPM LCT4\#6.}
\label{fig:wf-lct}
\end{figure}

\begin{figure}[h]
\centering
\vskip -0.2cm
\includegraphics[width=70mm]{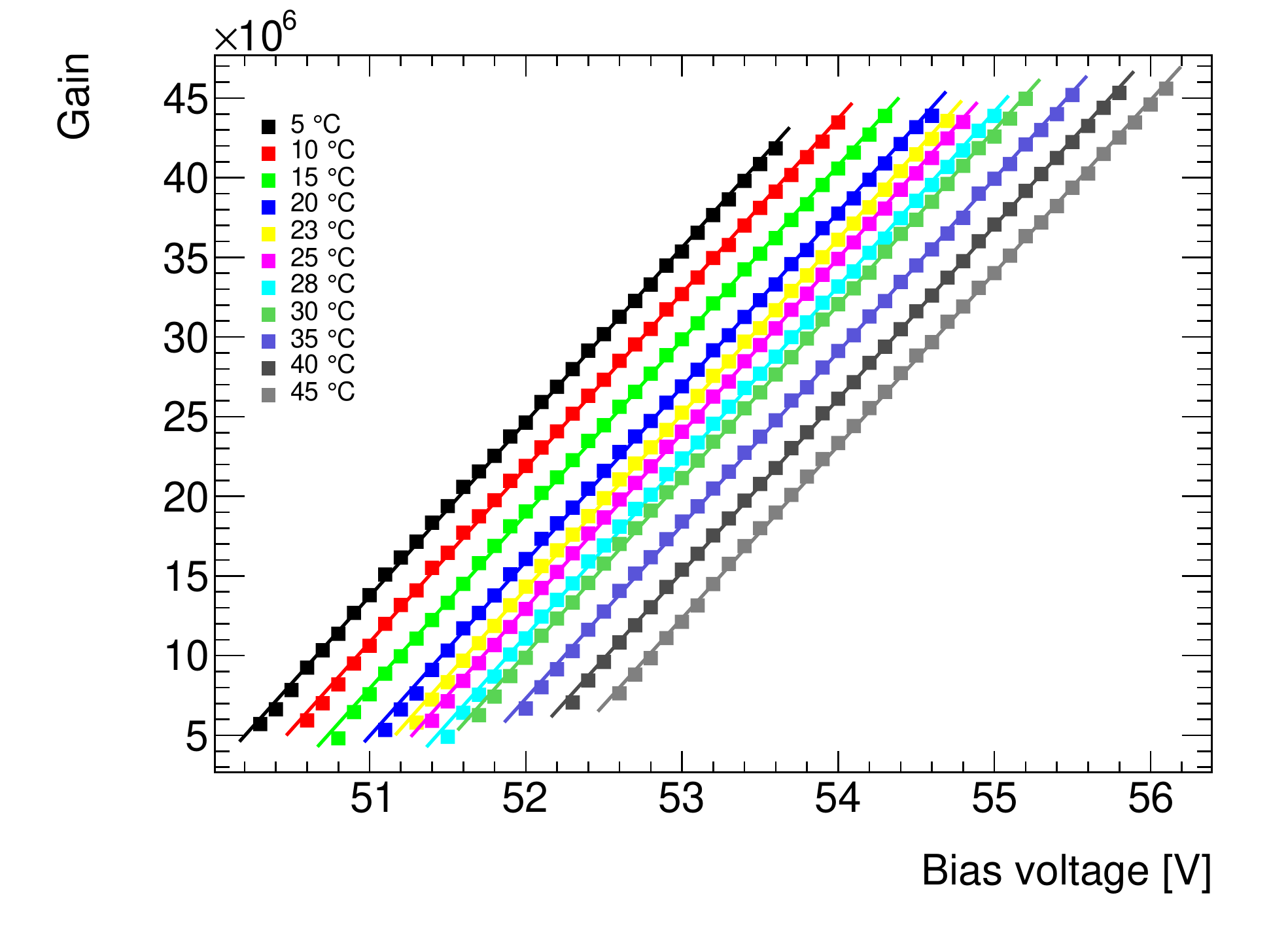}
\includegraphics[width=70mm]{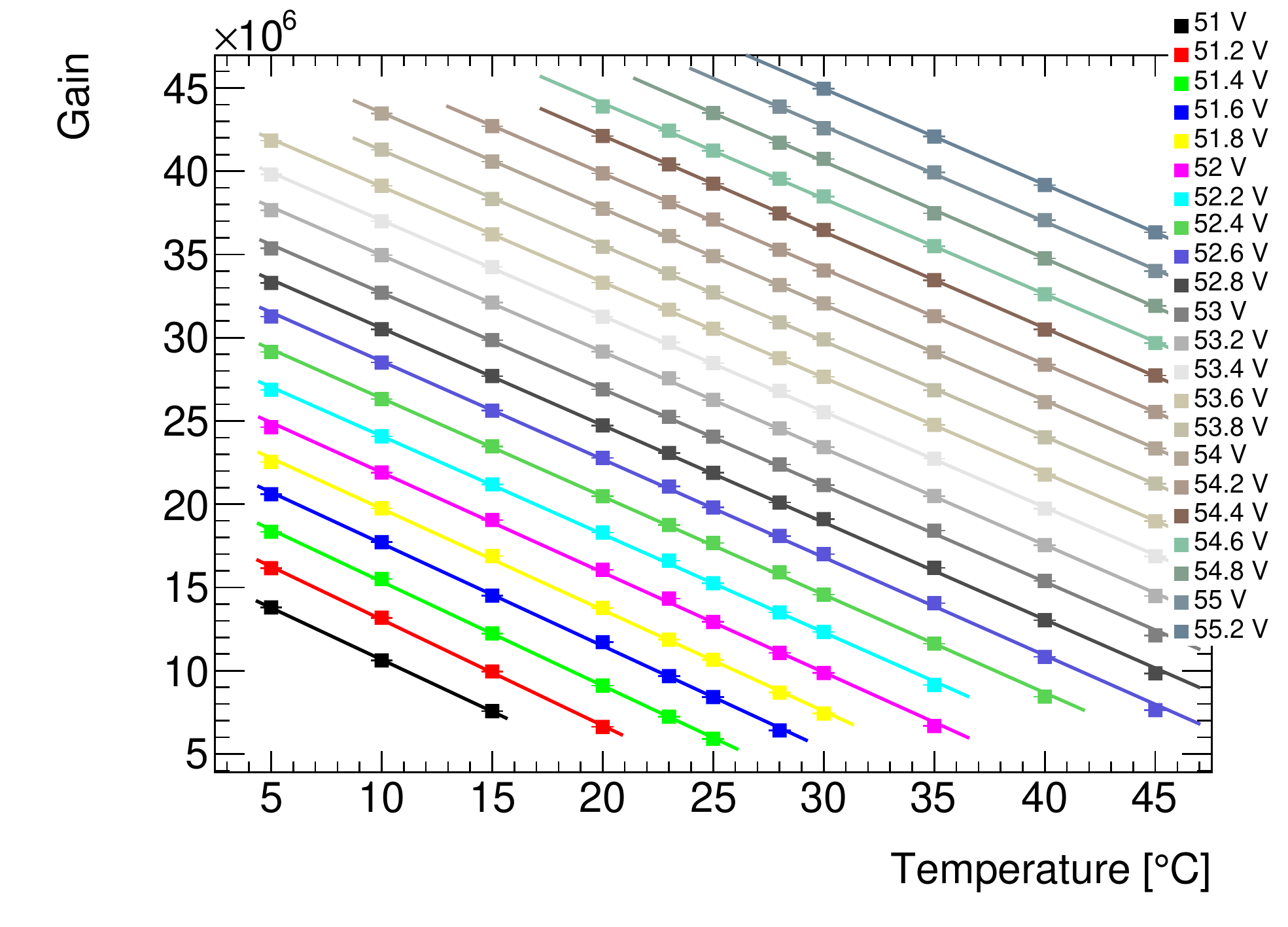}
\caption{Distributions of gain versus bias voltage for fixed temperatures (left) and gain versus temperature for fixed bias voltages (right) for SiPM LCT4\#6.}
\label{fig:dgdv-lct}
\end{figure}

We also tested two novel Hamamatsu SiPMs with trenches, LCT4\#6 and LCT4\#9. In these $\rm 1~mm \times 1~mm$ SiPMs with $\rm 50~\mu m$ pitch, the cross talk is reduced, leading in turn to reduced noise rates. The operation voltage at $25^\circ$C is $V_{bias} =54$ V. For example, Fig.~\ref{fig:wf-lct} (left) shows a typical waveform for  SiPMs LCT4\#6. The waveform  indicates  that noise and dark rate are at a low level. 
Figure~\ref{fig:wf-lct} (right) shows the corresponding photoelectron spectrum. Individual $pe$ peaks are well separated. A comparison with the $pe$ spectrum of SiPM B2 (Fig.~\ref{fig:pe1}) clearly demonstrates that for SiPM with trenches dark rates are substantially lower. The photoelectron peaks lie on a nearly uniform background. We perform a similar analysis as that for SiPM B2, measuring the gain versus bias voltage at fixed temperatures and then determining the gain versus temperature at fixed bias voltage.  Figures~\ref{fig:dgdv-lct} (left) and (right) depict the measurements with fit functions overlaid, respectively. The fit curves for the gain-versus-bias-voltage dependence at different temperatures are quasi parallel. Similarly, we observe nearly parallel fit curves for the gain-versus-temperature dependence at different \Vbias.

\begin{figure}[h]
\centering
\vskip -0.2cm
\includegraphics[width=50mm]{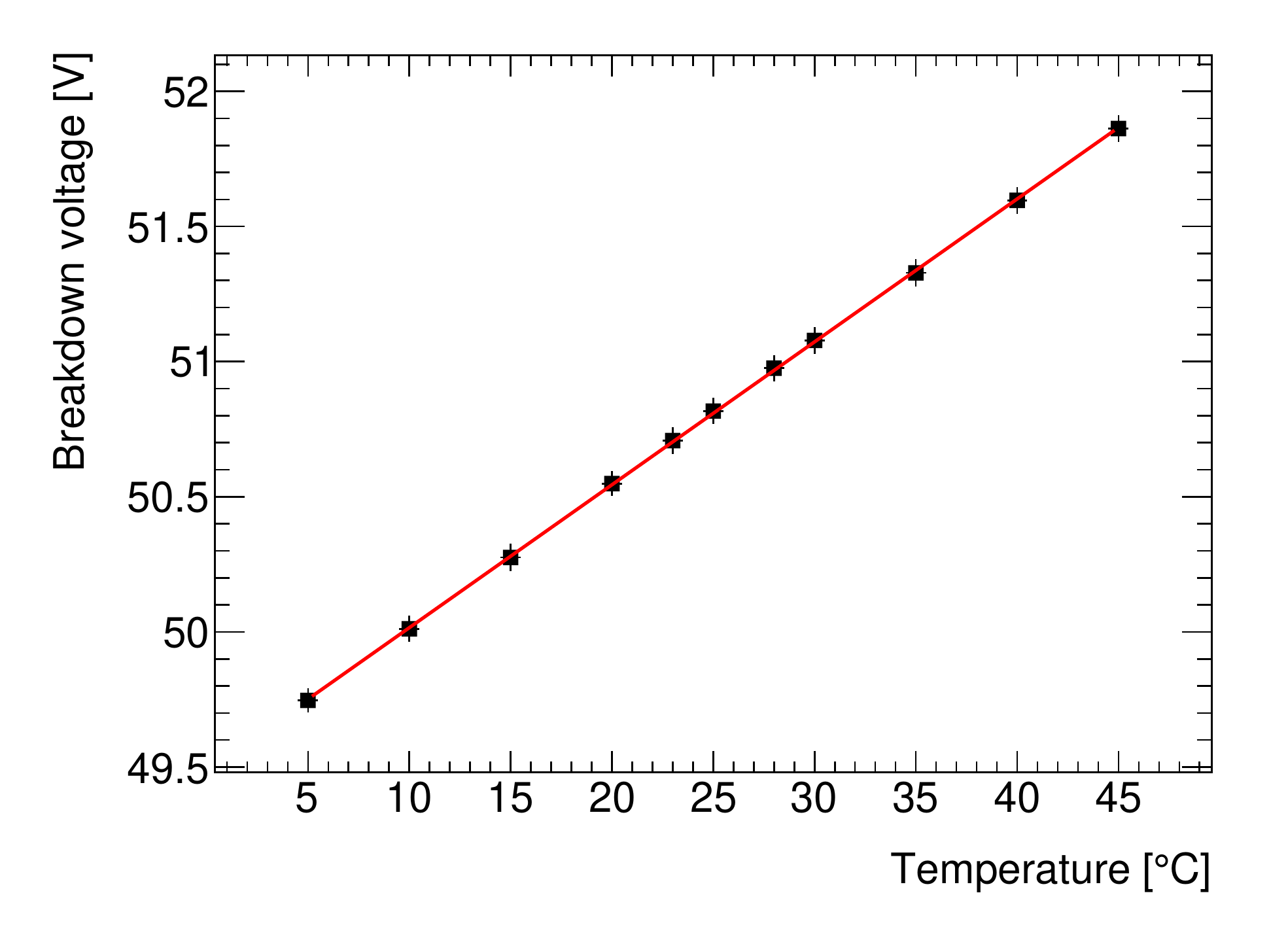}
\includegraphics[width=50mm]{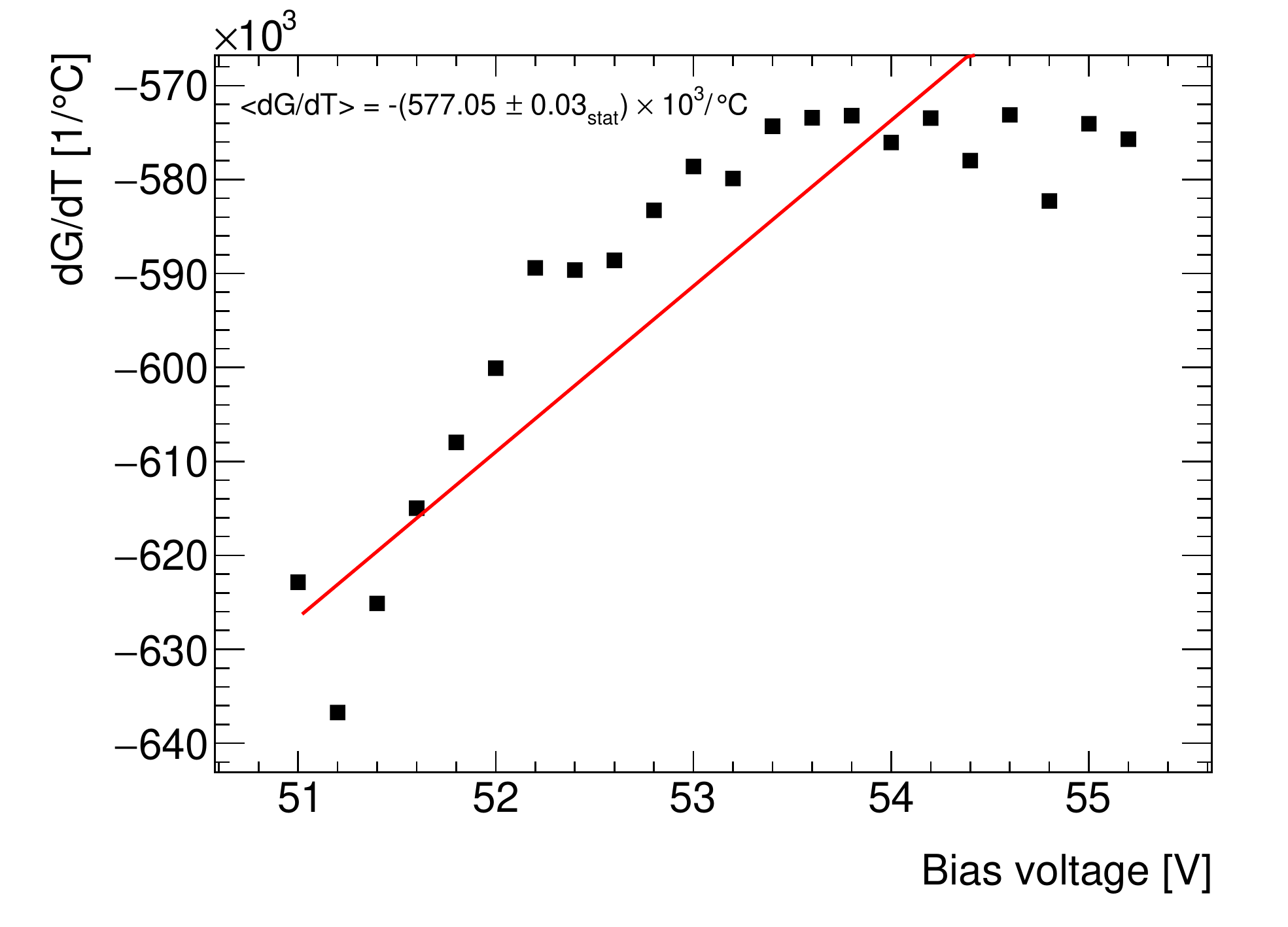}
\includegraphics[width=50mm]{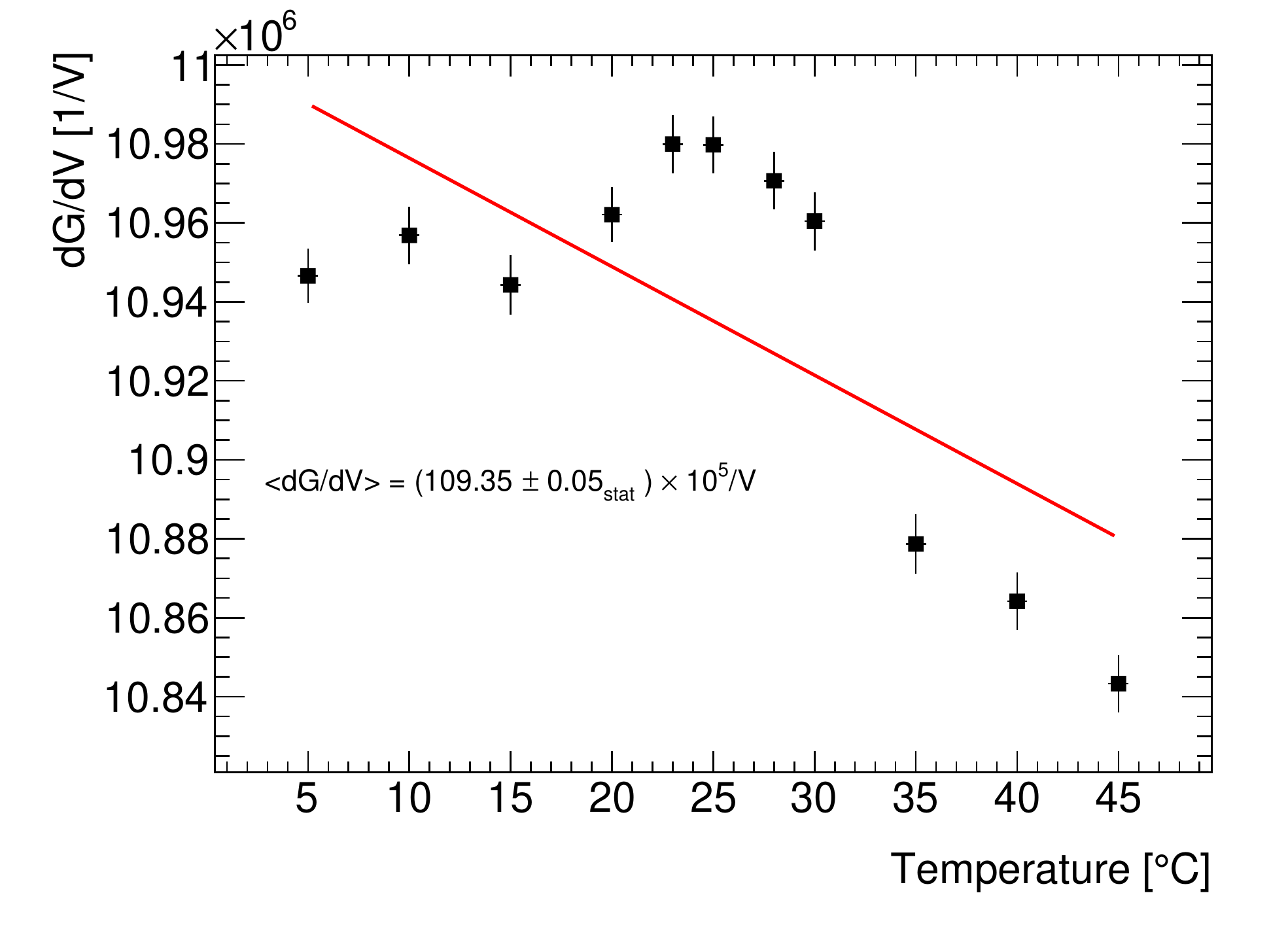}
\caption{Distributions of the break down voltage versus temperature (left), $dG/dV$ versus temperature (middle) and $dG/dT$ versus bias voltage (right) for SiPM LCT4\#6.}
\label{fig:break-lct}
\end{figure}

For LCT4\#6 at $25^\circ$C, we obtain $dG/dV= (10.935\pm 0.005) \times 10^6/$V. For the nominal bias voltage at $25^\circ$C,  we extract 
$dG/dT =- (0.57705 \pm 0.00003) \times 10^6/^\circ$C. Figures~\ref{fig:break-lct} {left), (middle) and (right) show the temperature dependence of $V_{break}$  and $dG/dV$ and the bias voltage dependence of $dG/dT$, respectively. The break-down voltage increases linearly with temperature. In the $5^\circ \rm C- 45^\circ$C temperature range, the capacitance $dG/dV$ is constant within $\pm 0.6\%$. The slope $dG/dT$ rises with bias voltage before leveling off around $53.5$~V. The entire increase is about $12\%$ in the $5^\circ \rm C - 45^\circ C$ temperature range. To determine $\overline{dV/dT}$, we extract ten individual values from different $dG/dV$ and $dG/dT$ combinations at each temperature, which we average. Figure~\ref{fig:dvdt-lct} (left) shows the  $\overline{dV/dT}$ values as a function of temperature, which is distributed rather uniformly. A fit yields the overall average value of $\langle dV/dT \rangle = 53.9\pm 0.5 ~\rm mV/^\circ C$. The corresponding measurements for LCT4\#9 are  $dG/dV= (10.802\pm 0.005) \times 10^6/$mV, $dG/dT =- (0.56281 \pm 0.00002) \times 10^6/^\circ$C and $\langle dV/dT \rangle = 54.0\pm 0.7 ~\rm mV/^\circ C$.
The $dV/dT$ results are 10\% lower than the specification of $60.0~\rm mV/^\circ C$ from Hamamatsu. The analytic solution yields a nearly linear dependence as shown in Fig.~\ref{fig:dvdt-lct} (right). The fit yields $a=(-1.52646 \pm 0.00009) \times 10^6$, $ b=17644 \pm 2$, $ c=(11.004 \pm 0.005) \times 10^6$, and $d=2749 \pm 192$ for  LCT4\#6  and $a=(-1.81156 \pm 0.00005) \times 10^6$, $ b=23134= \pm 1$, $ c=( 10.774\pm 0.005) \times 10^6$, and $d=1130 \pm 170$ for  LCT4\#9.

\begin{figure}[h]
\centering
\vskip -0.2cm
\includegraphics[width=70mm]{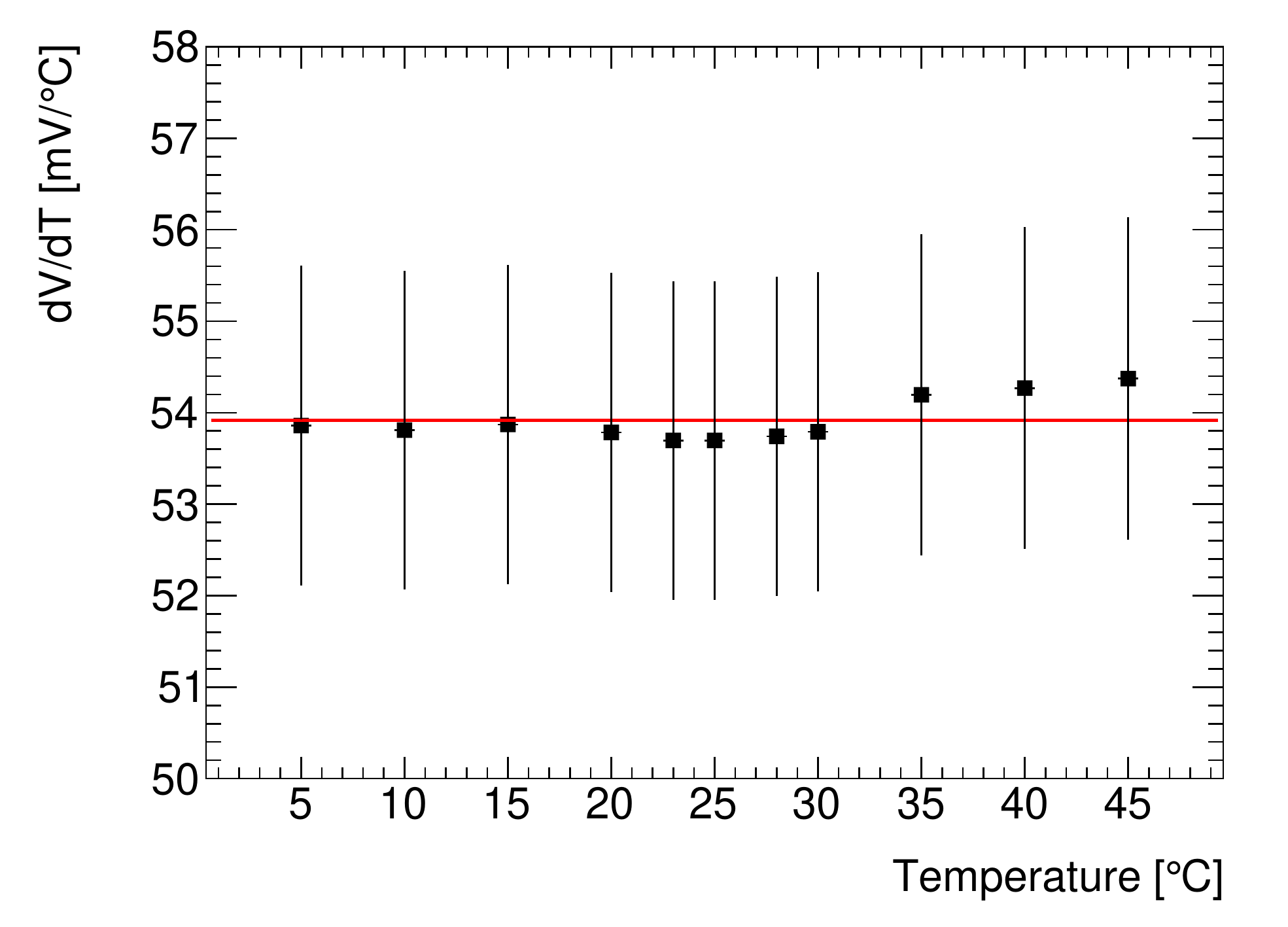}
\includegraphics[width=70mm]{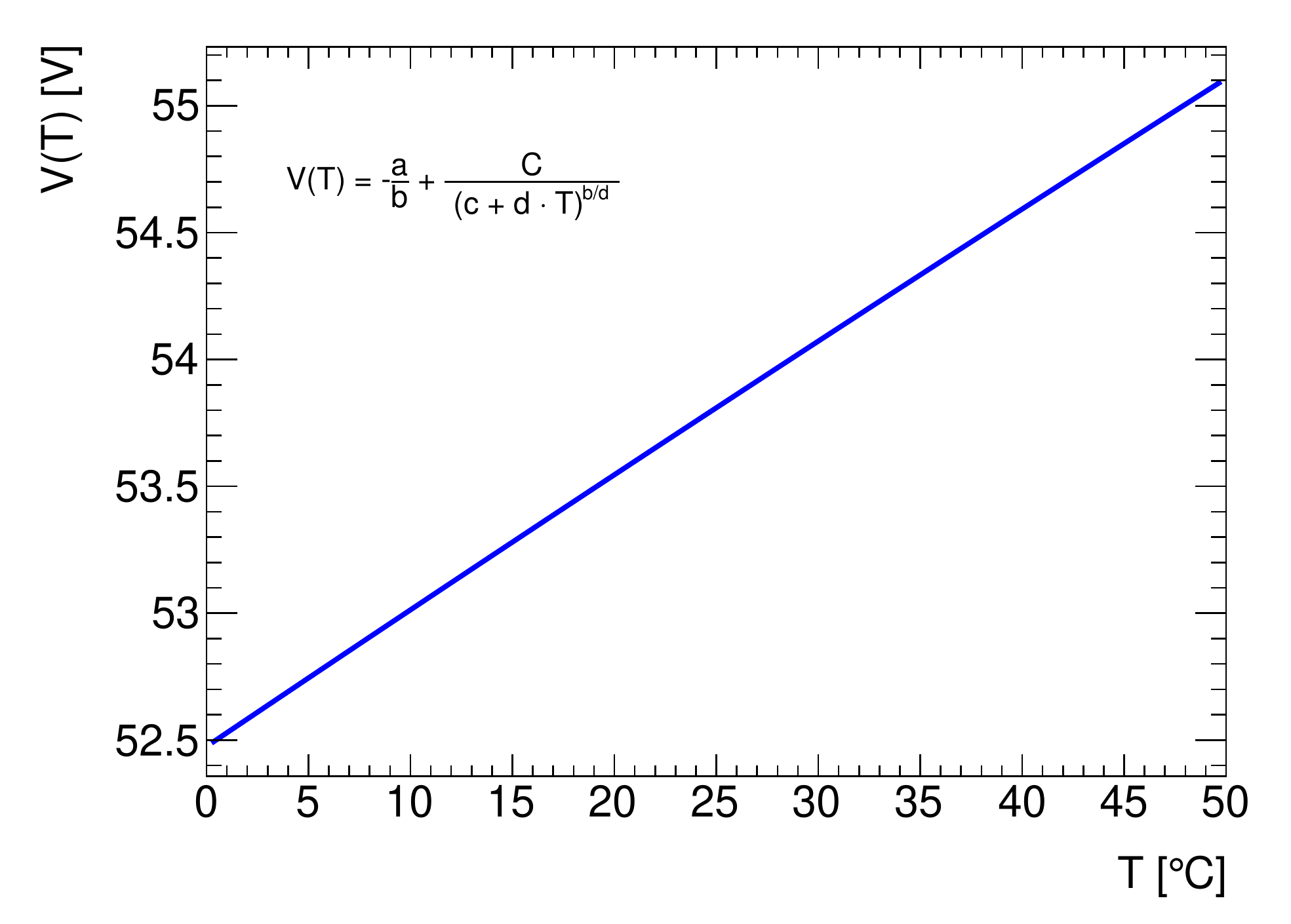}
\caption{Distributions of $\overline{dV/dT}$ versus temperature (left) and $V(T)$ versus temperature (right) for SiPM LCT4\#6.} 
\label{fig:dvdt-lct}
\end{figure}

\begin{figure}[h]
\centering
\vskip -0.2cm
\includegraphics[width=70mm]{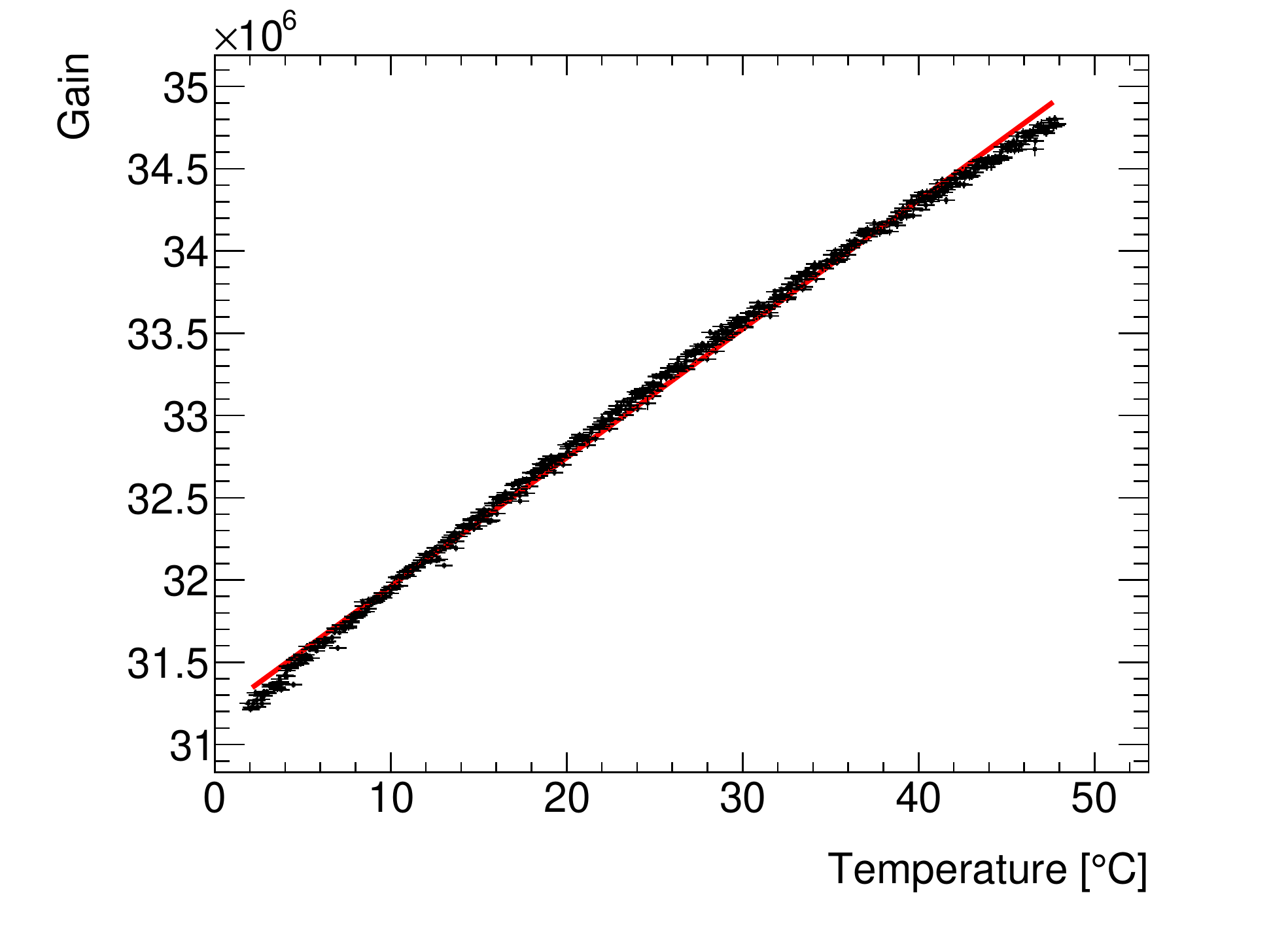}
\includegraphics[width=70mm]{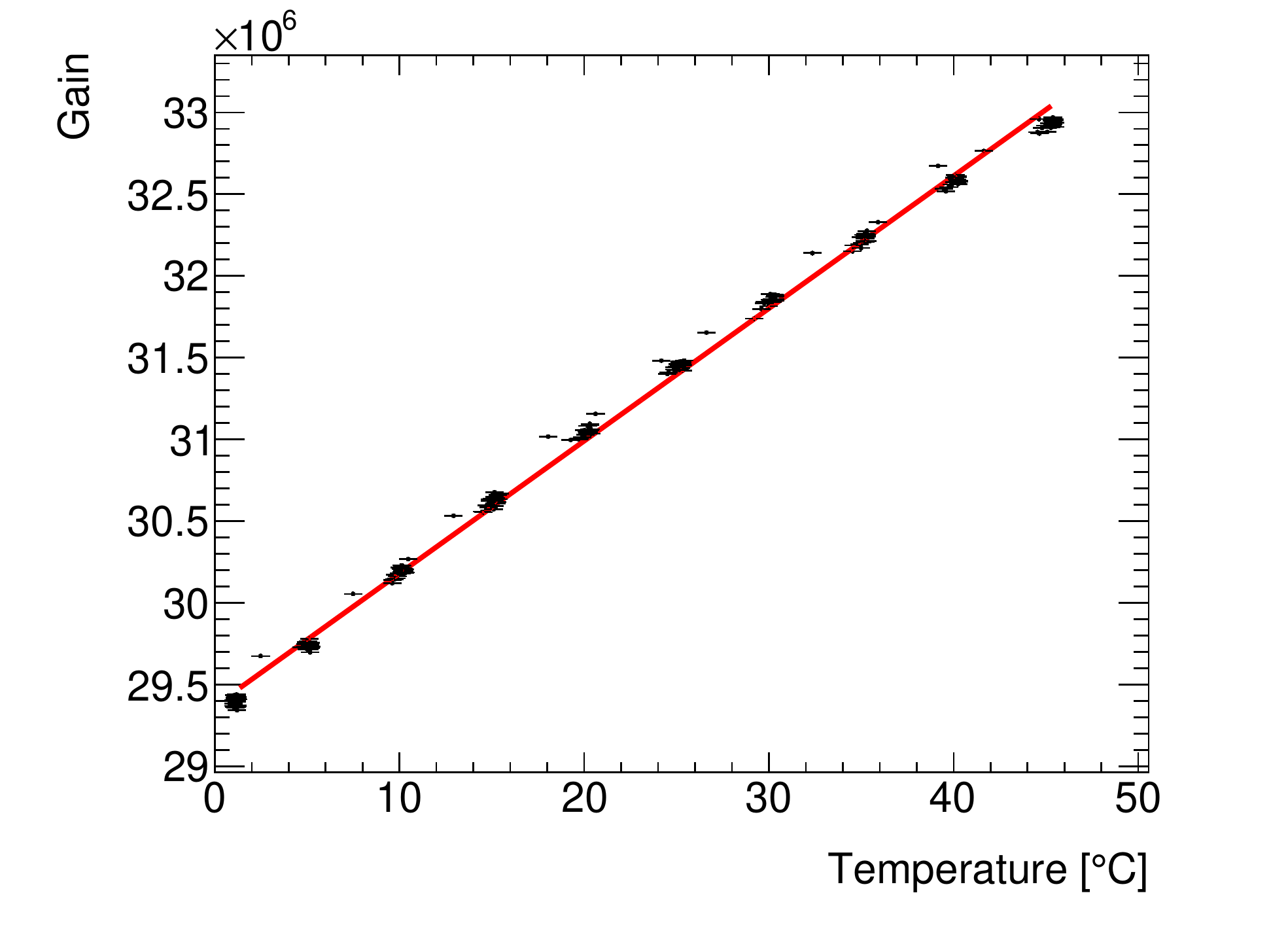}
\caption{Results of gain stabilization for LCT4\#6 with a continuous temperature scan (left) and LCT4\#9 (right) with a step-wise temperature scan .} 
\label{fig:stabilization-lct}
\end{figure}

We performed the gain stabilization with $dV/dT=60~mV/^\circ\rm  C$ as given by Hamamatsu. Figure~\ref{fig:stabilization-lct} shows the gain-versus-temperature dependence after stabilization for SiPM LCT4\#6 (left) using a continuous temperature scan and for LCT4\#9 using a stepwise temperature scan (right). The gain is rising with temperature. Up to $\sim35^\circ$C, the rise is linear. Above $\sim40^\circ$C, a deviation from linearity is visible. In the $20^\circ \rm C - 30^\circ C$ temperature range, the deviation from uniformity is $\pm 1.2\%$.  This is a somewhat larger than our requirement.  Thus, we tested the gain stabilization of these SiPMs with the correct slope in a new study at CERN in February 2016.

\section{Measurements of KETEK SiPMs}

Besides Hamamatsu SiPMs, we tested sensors from KETEK and CPTA. The CPTA SiPMs ($\rm 1~mm \times 1~mm$ with a pitch of $\rm 40~\mu m$) were glued to a wavelength-shifting fiber inserted into a groove milled into a 3~mm thick and $\rm 3~cm \times 3~cm$ wide scintillator tile. This made the light injection challenging. We cleared the wrapping near the SiPM to inject light close to the sensor. Since the photoelectron spectra have larger background than those of other SiPMs, they are much harder to fit. Thus, the analysis is still ongoing. 

The KETEK SiPM, W12, is $\rm 3~mm \times 3~mm$ experimental sensors with a pitch of  $\rm 20~\mu m$. The nominal operation voltage is $V_{bias}=28$ V at $25^\circ$C. We perform a similar analysis as before. Figure~\ref{fig:dgdv-w12} (left) shows the gain-versus-bias-voltage dependence at fixed temperatures. The fits yield nearly parallel lines. Figure~\ref{fig:dgdv-w12} (right) shows the gain-versus-temperature dependence at fixed bias voltage. Some fit curves are slightly non-parallel. This reflects a larger overall systematic uncertainty. 
Figure~\ref{fig:break-w12} (left) and (middle) show the breakdown voltage and $dG/dV$ versus temperature, respectively. The latter variable rises linearly with temperature. The deviation from constant capacitance is $5\%$ from  $5^\circ\rm  C$ to $45^\circ$C. Figure~\ref{fig:break-w12} (right) shows $dG/dT$ versus bias voltage, which increases with bias voltage. The variation from low to high bias voltages is 23\%.
At $\rm 25^\circ C$, we measure $ dG/dV= (2.9293\pm 0.0004)\times10^6/$V and $ dG/dT =-(0.25336 \pm 0.0004)\times 10^6/^\circ$C. Again, we average ten $dV/dT$ points at each temperature and fit the resulting gain distribution shown in Fig.~\ref{fig:dvdt-w12} (left). The fit yields
 $\langle  dV/dT\rangle   = 17.2\pm 0.4\rm ~ mV/^\circ C$. This is somewhat smaller than the measurement of  $\langle dV/dT\rangle = 21.29\pm 0.08\rm ~ mV/^\circ C$ obtained in a previous study. Figure~\ref{fig:dvdt-w12} (right) shows the analytical solution for $V(T)$. The curve is obtained with  $  a=-226823\pm 32$,  $b=6297.4\pm 1.7$, $c=(2.9231 \pm 0.008) \times 10^6$ and $d=3792 \pm 389$.

\begin{figure}[h]
\centering
\vskip -0.2cm
\includegraphics[width=70mm]{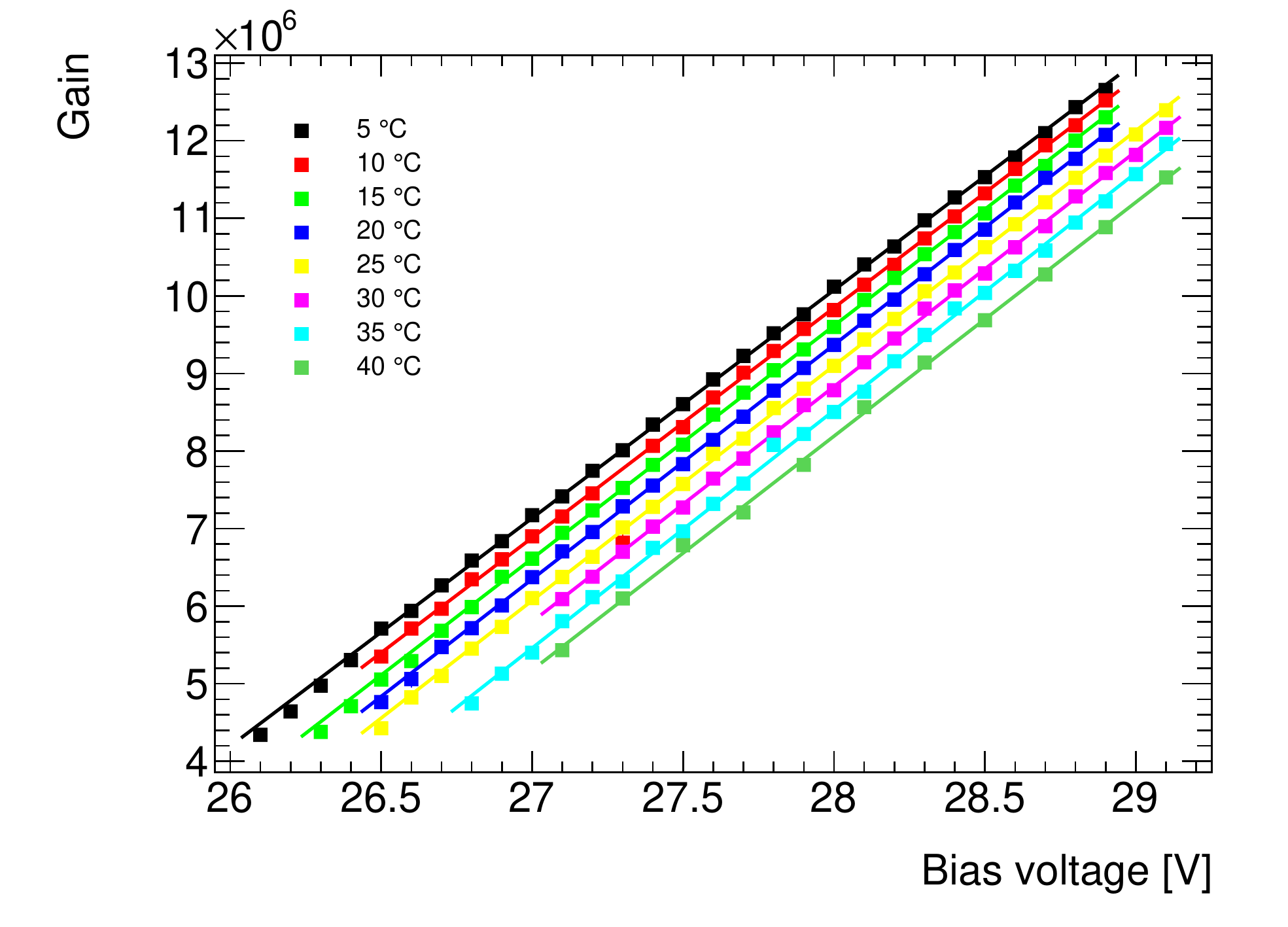}
\includegraphics[width=70mm]{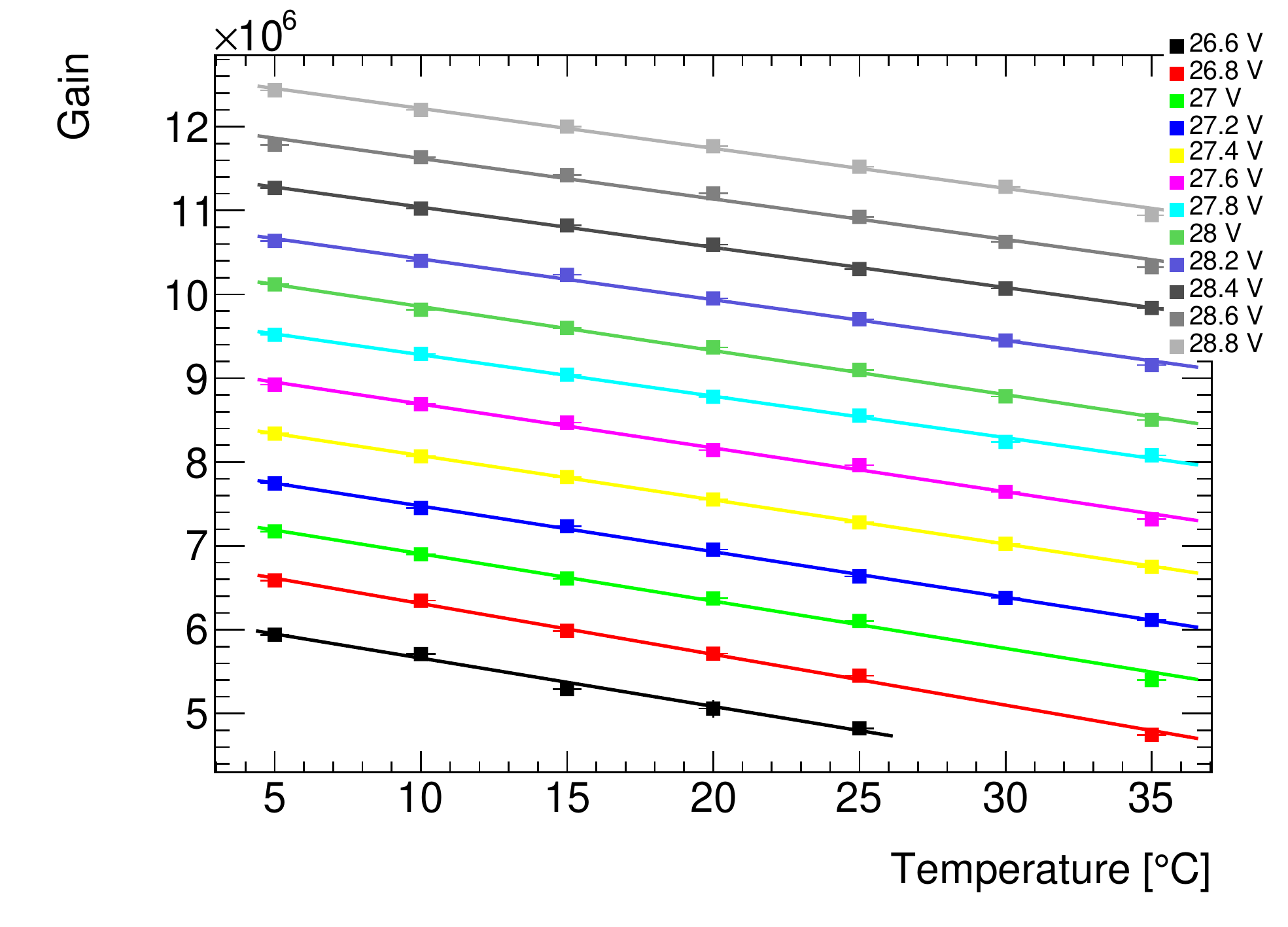}
\caption{Distributions of gain versus bias voltage for fixed temperatures (left) and gain versus temperature for fixed bias voltages (right) for SiPM  W12 from KETEK.}
\label{fig:dgdv-w12}
\end{figure}
 
\begin{figure}[h]
\centering
\vskip -0.2cm
\includegraphics[width=50mm]{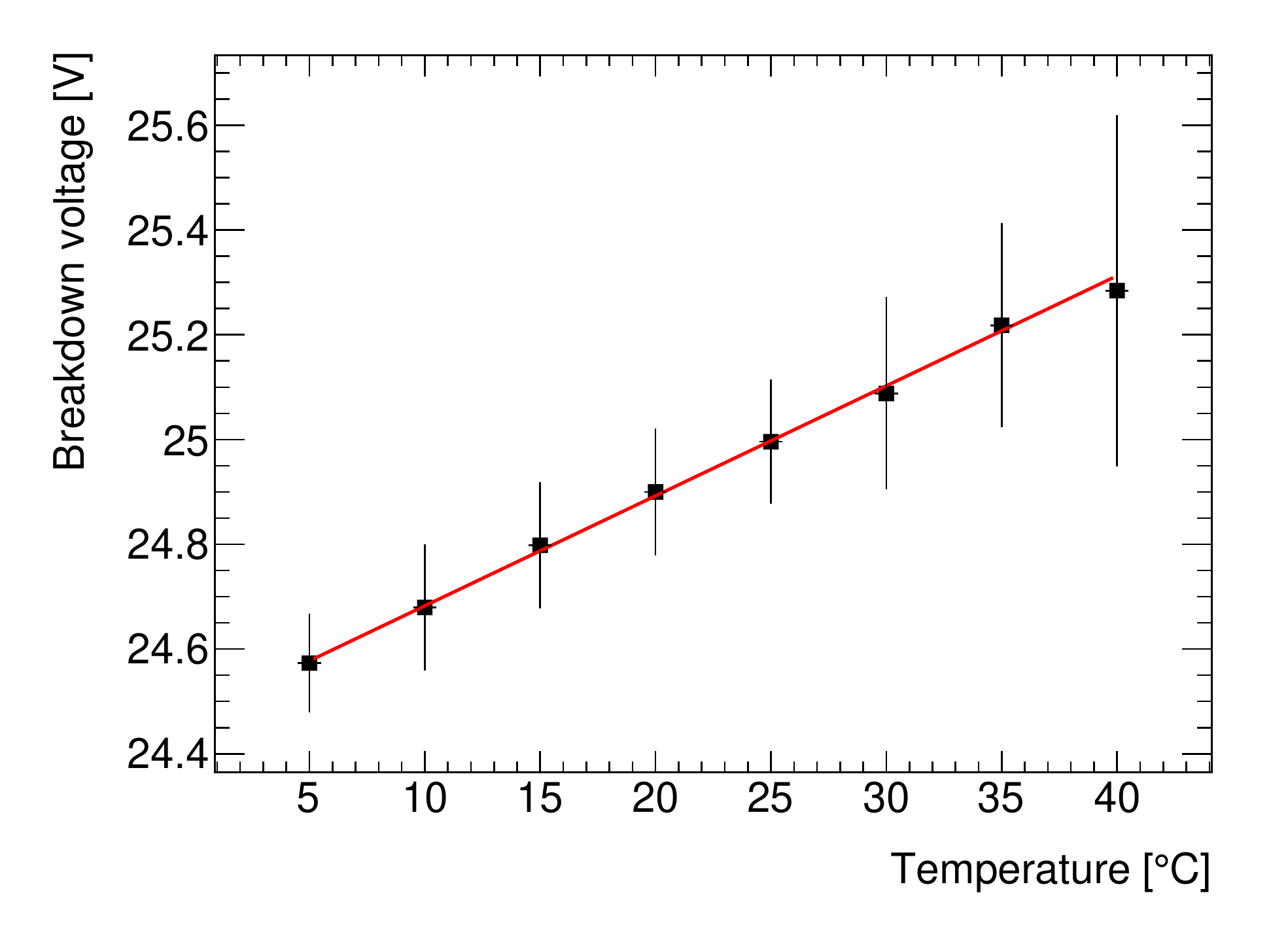}
\includegraphics[width=50mm]{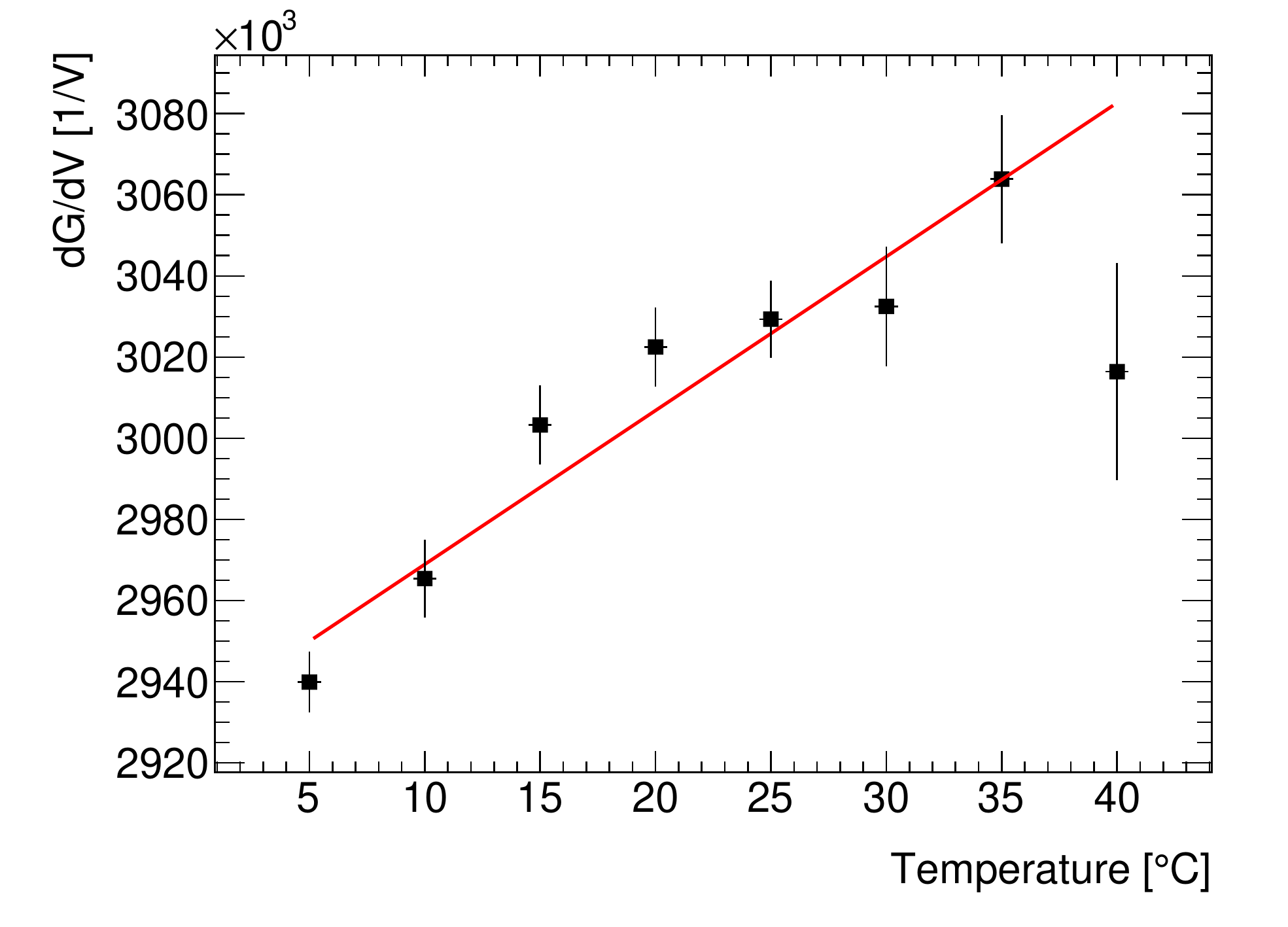}
\includegraphics[width=50mm]{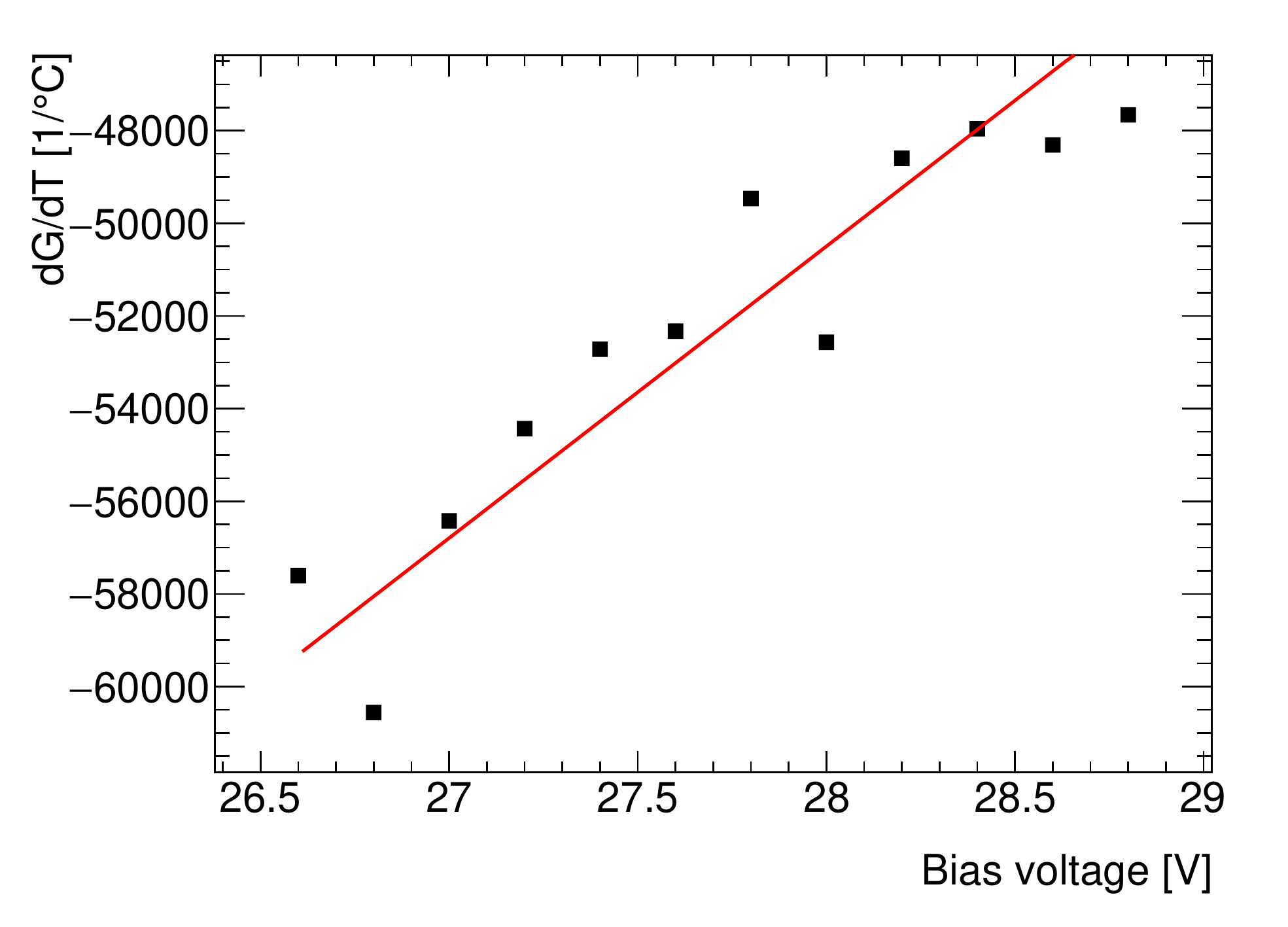}
\caption{Distributions of the break down voltage versus temperature (left), $dG/dV$ versus temperature (middle) and $dG/dT$ versus bias voltage (right) for SiPM W12 from KETEK.}
\label{fig:break-w12}
\end{figure}

\begin{figure}[h]
\centering
\vskip -0.2cm
\includegraphics[width=70mm]{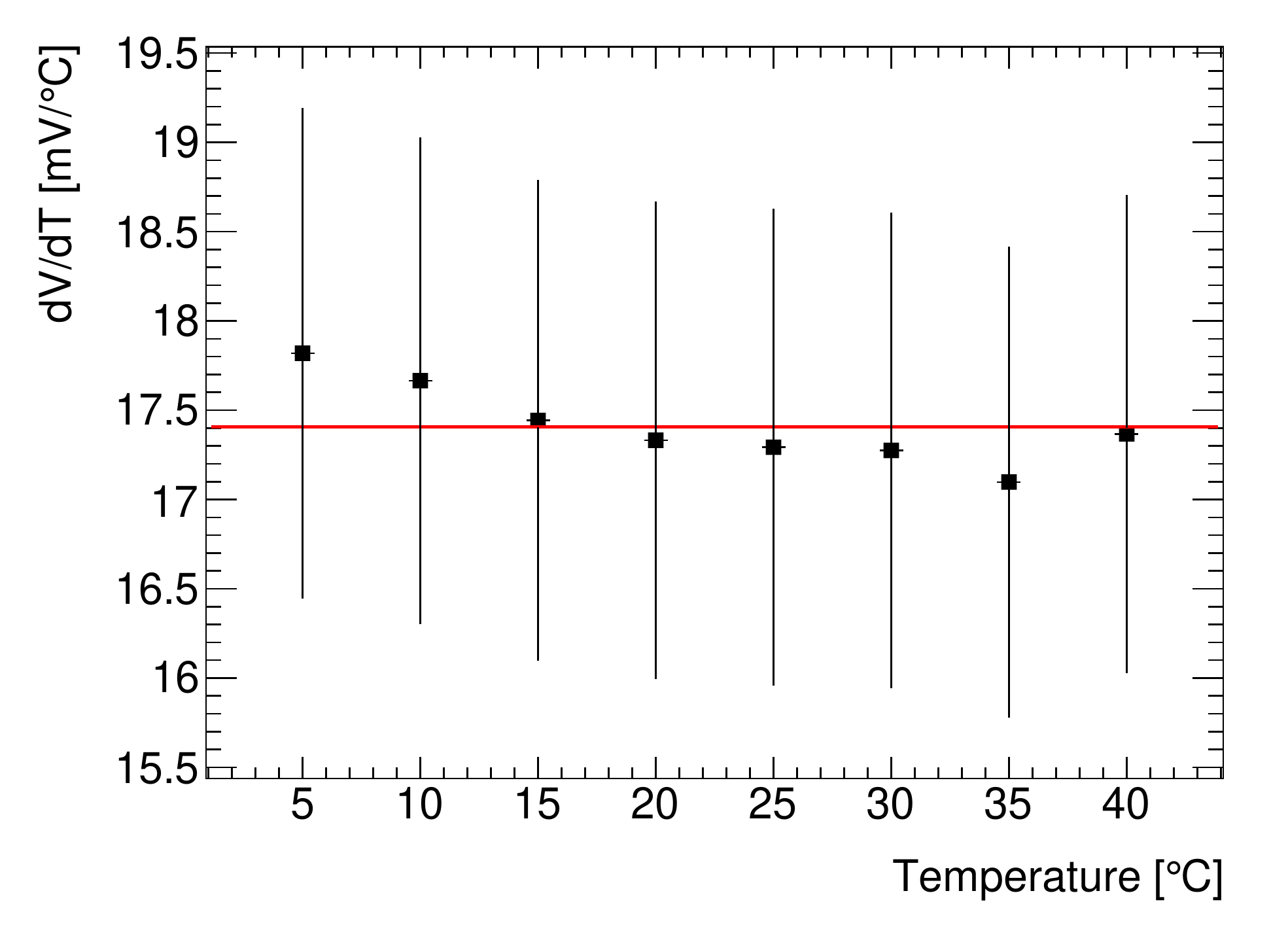}
\includegraphics[width=70mm]{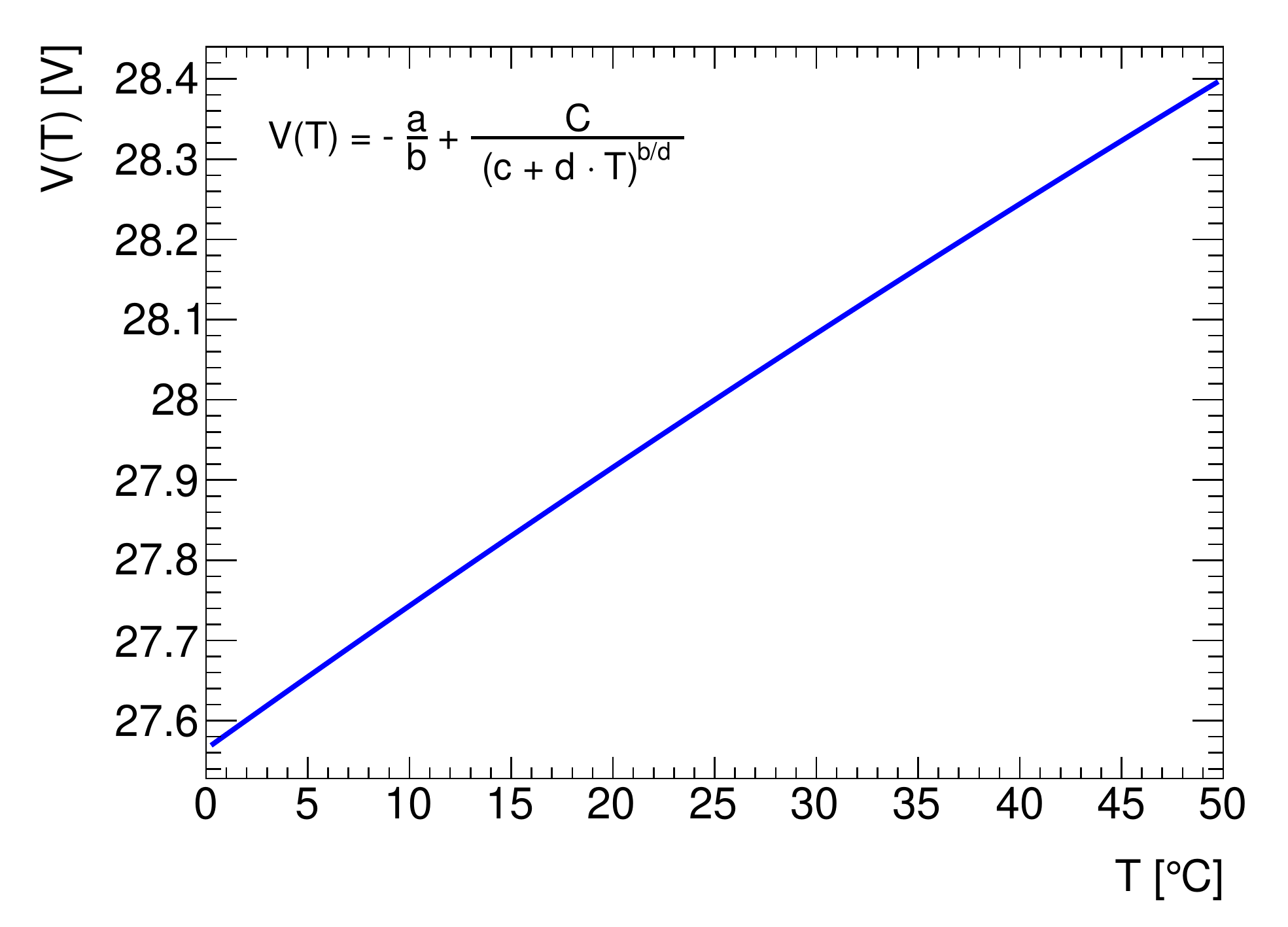}
\caption{Distributions of $\overline{dV/dT}$ versus temperature (left) and $V(T)$ versus temperature (right) for SiPM  W12 from KETEK.} 
\label{fig:dvdt-w12}
\end{figure}
 
Since we had not extracted $\langle dV/dT \rangle$ before the gain stabilization study, we used the slope of $21~\rm mV/^\circ \rm  C$ from a previous study\cite{cvach}.  Figure~\ref{fig:ramp-w12} shows the gain versus temperature dependence after stabilization. The distribution shows a linear increase with temperature to about $35^\circ$C. In the $0^\circ \rm - 35^\circ$C temperature range, the overcompensation is about $\pm 2.8\%$ yielding a deviation from stability of $\pm 0.8\%$ in the $20^\circ \rm C - 30^\circ$ C temperature range.  Above  $35^\circ$C, the gain starts dropping. According to the KETEK data sheet, the SiPMs should be operated at temperatures between $-30^\circ$C to  $+40^\circ$C.  So data above  $+35^\circ$C are not stable. Since the deviation from uniformity lies outside our specification, we remeasured the gain stabilization with the correct $\langle dV/dT \rangle$ value in the second test at CERN in February 2016.

\begin{figure}[h]
\centering
\vskip -0.2cm
\includegraphics[width=70mm]{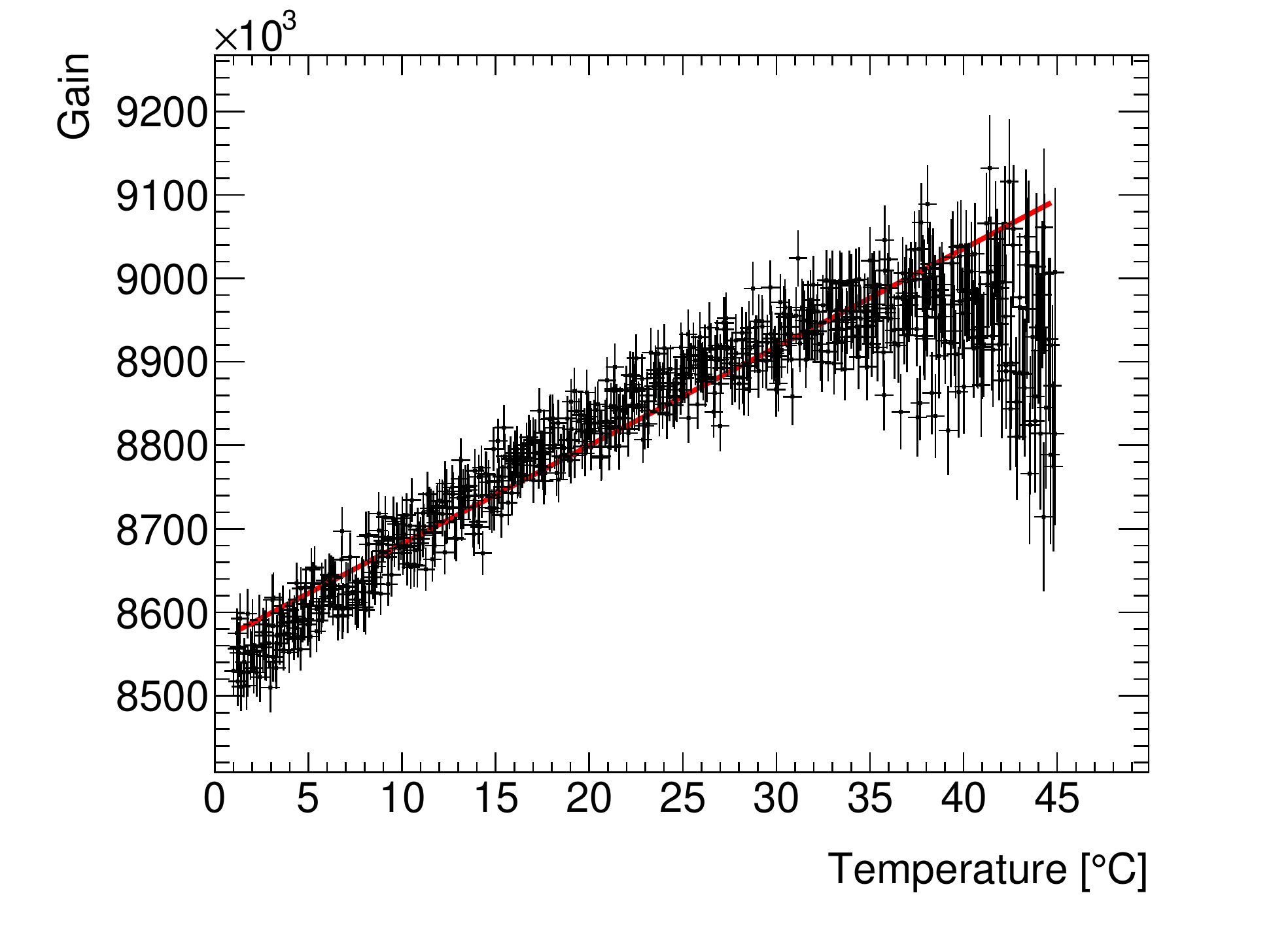}
\caption{Distributions of gain versus temperature after stabilization for continuous ramping of the temperature for SiPM  W12 from KETEK.} 
\label{fig:ramp-w12}
\end{figure}

\section{Studies of Afterpulsing}

The SiPM pulse has a very fast rise time and a rather long decay time. The decay time depends on the intrinsic properties of the SiPM (quenching resistor, capacitance) and properties of the preamplifier. The quenching resistor terminates the avalanche formation to get the detector ready for the next signal. In this process, however, a new avalanche may be triggered called afterpulse that affects the shape of the waveform on the decay time. The probability for afterpulses depends on the internal properties of the SiPM. Since afterpulses are delayed with respect to the original signal, their charge is only partially integrated over. This leads to photoelectron spectra with decreased resolution. In order to measure the size of afterpulsing, we compare photoelectron spectra obtained with two different methods. First, we determine the photoelectron spectrum from the measured total charge  $Q_{tot}$ by integrating the waveform over a fixed time window. Second, we extract the photoelectron spectrum from the magnitude of the waveform peak $A_{peak}$. Figure~\ref{fig:scatter-B2} (left) shows the scatter plot of $Q_{tot}$ versus $A_{peak}$ for SiPM B2 at $25^\circ$C for the nominal bias voltage of $74.9$ V. The signal without afterpulsing lies on the diagonal. This is the main contribution at this bias voltage. Waveforms with afterpulses shift $Q_{tot}$ upwards due to the additional charge. 
Thus, we define a separation line shown by the dashed line in Figure~\ref{fig:scatter-B2} (left). The slope is obtained from the separation of the second and third photoelectron peaks in two dimensions, $\Delta y/ \Delta x$. To achieve the best separation between the two regions, the position of the line is chosen at the minimum value between the two regions as shown in Fig.~\ref{fig:scatter-B2} (right). To determine the fraction of afterpulsing, we count all events above the dashed line and normalize them to the total number of events. In order to test if the determination of $dV/dT$ depends on afterpulsing, we redo the analysis for all events and those above the dashed line. Figures~\ref{fig:red-afterpulsing} (top left) and (top right) show $dG/dV$ versus temperature for all waveforms and those with reduced afterpulsing. Figures~\ref{fig:red-afterpulsing} (bottom left) and (bottom right) show the corresponding plots for $dG/dT$ versus bias voltage. The $dG/dV$ and $dG/dT$ distributions for the two samples look similar and the slopes both for $dG/dV$ for $dG/dT$ are the same within errors.

\begin{figure}[h]
\centering
\vskip -0.2cm
\includegraphics[width=70mm]{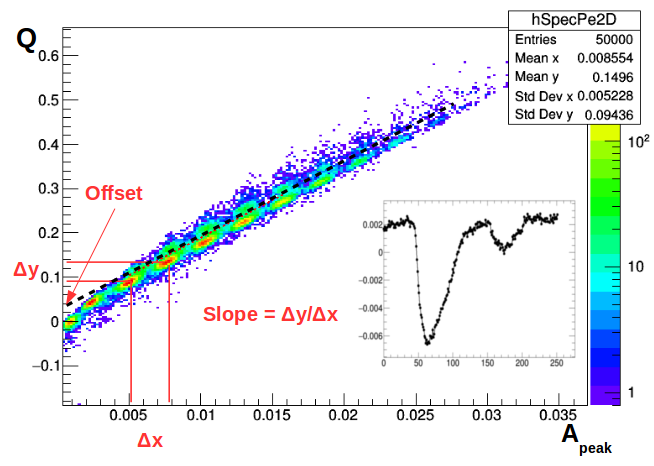}
\includegraphics[width=70mm]{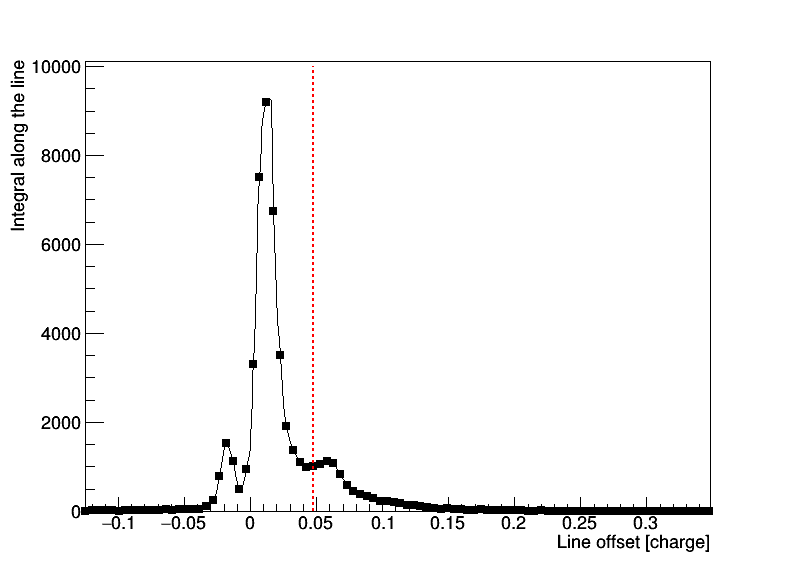}
\caption{Scatter plot of photoelectron spectra extracted from $Q_{tot}$ versus those extracted from $A_{peak}$ for SiPM B2 (left) and the one-dimensional projection onto the an axis orthogonal to the diagonal (right). The dashed line indicates the separation between waveforms with afterpulsing to those without.} 
\label{fig:scatter-B2}
\end{figure}

  \begin{figure}[h]
\centering
\vskip -0.2cm
\includegraphics[width=70mm]{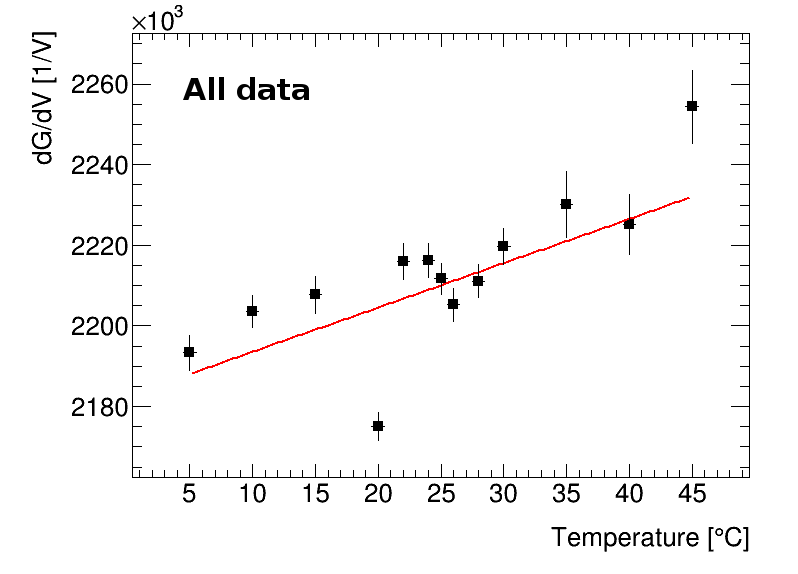}
\includegraphics[width=70mm]{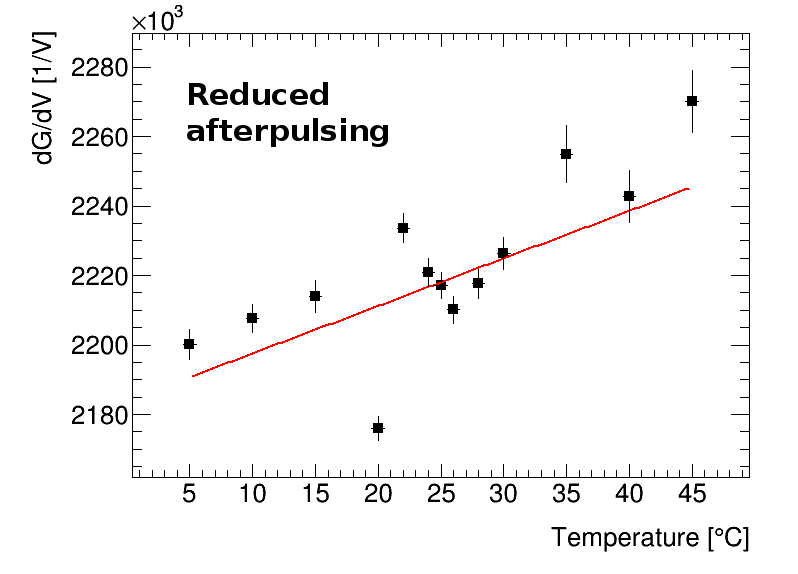}
\includegraphics[width=70mm]{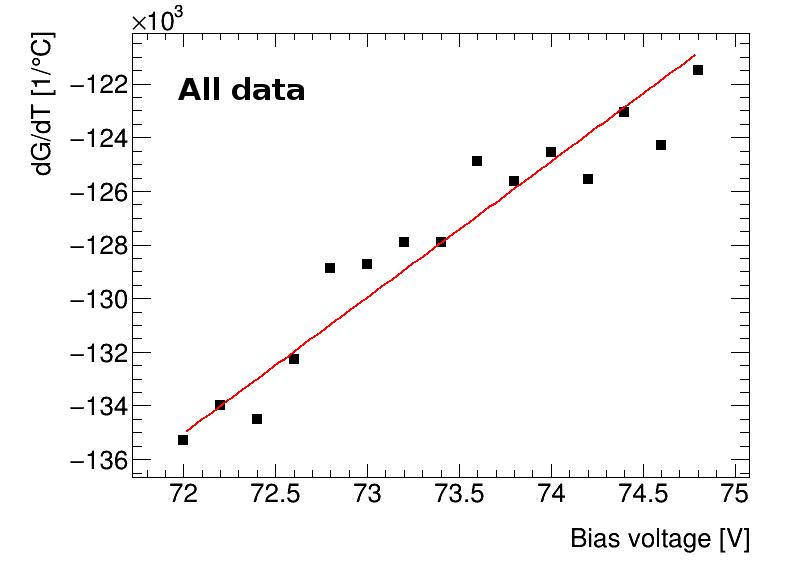}
\includegraphics[width=70mm]{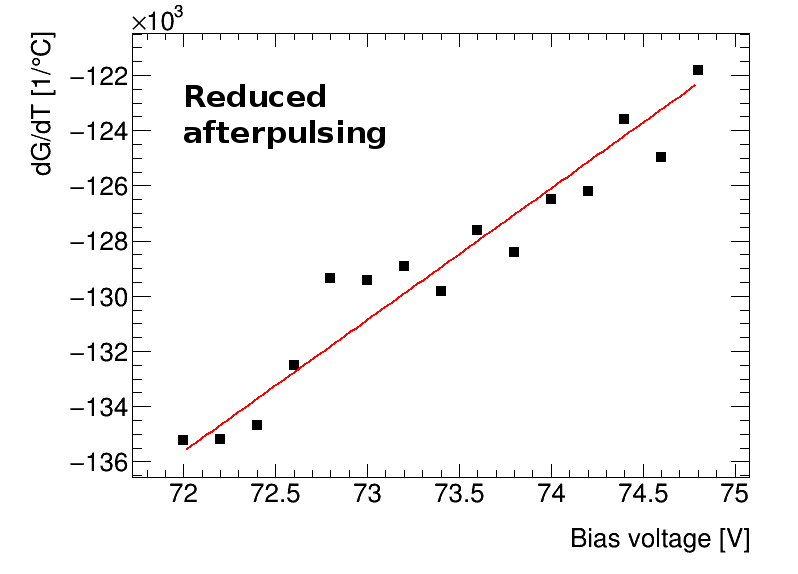}
\caption{The $dG/dV$-versus-temperature distributions for  all events (top left) and events with reduced afterpulsing (top right) and $dG/dT$-versus-$V$ distributions for  all events (bottom left) and events with reduced afterpulsing (bottom right) for SiPM B2.} 
\label{fig:red-afterpulsing}
\end{figure}

\begin{figure}[h]
\centering
\vskip -0.2cm
\includegraphics[width=70mm]{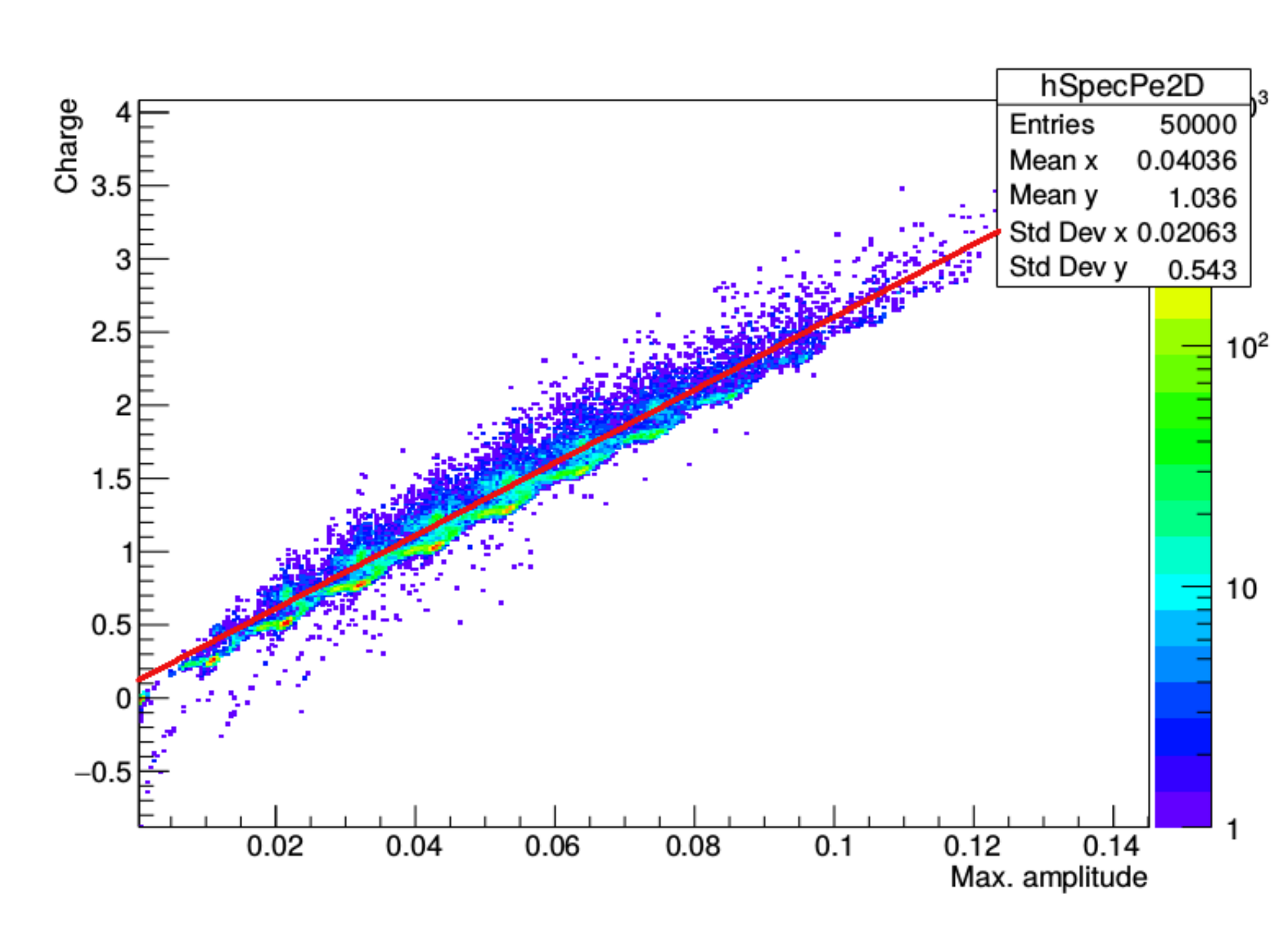}
\caption{Scatter plot of photoelectron spectra extracted from $Q_{tot}$ versus those extracted from $A_{peak}$ for SiPM LCT4\#6. The red (solid) line indicates the separation between waveforms with afterpulsing to those without.} 
\label{fig:scatter-LCT}
\end{figure}

 \begin{figure}[h]
\centering
\vskip -0.2cm
\includegraphics[width=70mm]{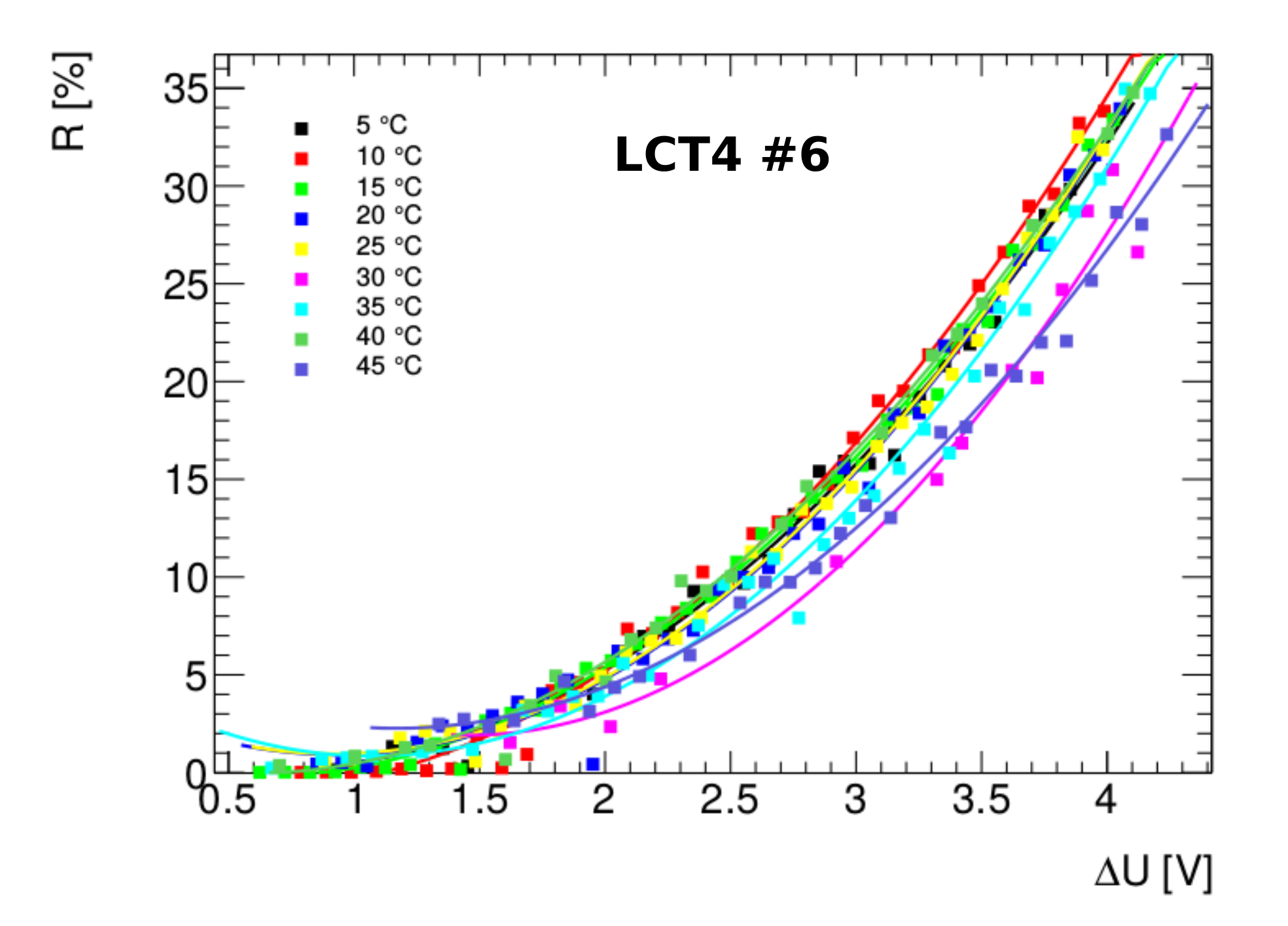}
\includegraphics[width=70mm]{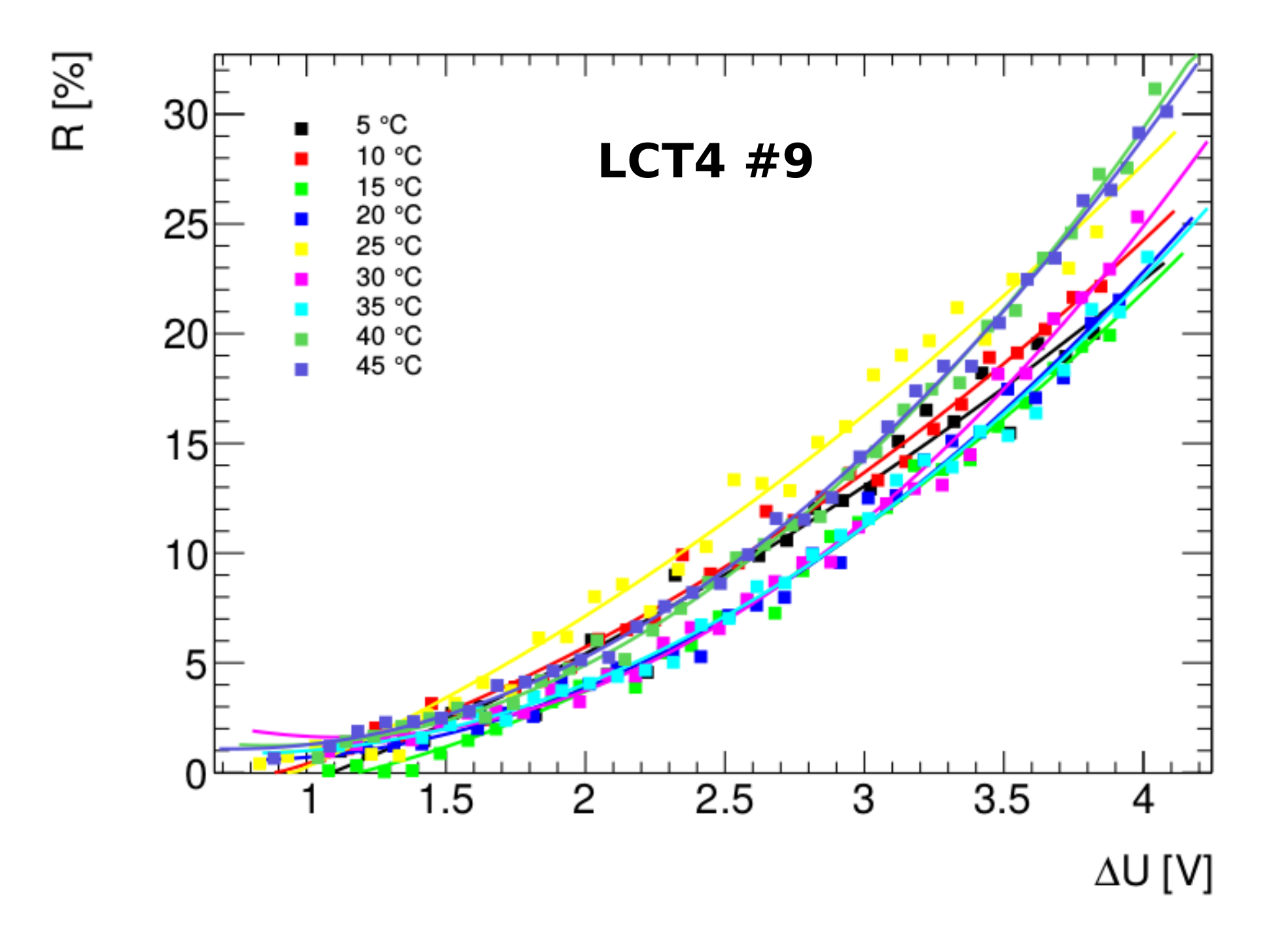}
\caption{The distributions of the afterpulsing fraction R versus overvoltage for SiPM LCT4\#6 (left) and LCT4\#9 (right) for different temperatures.} 
\label{fig:R}
\end{figure}

To determine the afterpulsing fraction for LCT4 SiPMs, we analyze scatter plots of photoelectron spectra  for $Q_{tot}$ versus $A_{peak}$ for different temperatures and bias voltages. Figure~\ref{fig:scatter-LCT} shows the scatter plot for LCT4\#6 at $25^\circ$C at the nominal bias voltage.
We define the afterpulsing fraction R as the number of events above the red line to the number of all events. For different temperatures, we plot $R$ as a function of the overvoltage $\Delta U =V_{bias}- V_{break}$. Figure~\ref{fig:R} shows the results for SiPM LCT4\#6 (left) and LCT4\#9 (right). An increase with overvoltage is visible that can be fitted with a second-order polynomial. However, $R$ shows no significant dependence on temperature.
The observed spread is consistent with the systematic uncertainty of the measurement procedure.

\section{Conclusion and Outlook}

In the framework of AIDA2020, we started to conduct gain stabilization studies of more SiPMs in the climate chamber at CERN with an improved readout system using a 12-bit digital oscilloscope controlled by LabView.  We set up a readout of two SiPMs in parallel, one  attached to
a voltage-operational  preamplifier and the other to a current-operational preamplifier. The current-operational preamplifier became unstable during the gain stabilization tests, particularly after operation at higher temperatures producing useless results that had to be discarded.  
Thus, out of the eight SiPMs we wanted to study, we could test gain stabilization only for five of them.
Two are novel SiPMs from Hamamatsu with trenches (LCT4\#6, LCT4\#9), one is special SiPM with $\rm 15~\mu m$ pitch from Hamamatsu (B2), one comes from KETEK (W12) and the fifth is a CPTA sensor (857). 
For SiPM B2, we achieved excellent gain stabilization in entire temperature range ($\rm 5^\circ C-45^\circ$C).  The deviation from stable gain is less than $1\%$ in the entire temperature range. For SiPMs LCT4\#6 and LCT4\#9, we overcorrected the bias voltage by $+6.1 \rm mV/^\circ \rm C$ since we used $dV/dT=60 \rm ~mV/^\circ C$  specified by Hamamatsu instead of $53.9 \rm ~mV/^\circ C$ measured in this study. This yields a deviation from stability  of $\pm 1.2\%$ in the $\rm 20^\circ C-30^\circ$C temperature range, which is more than a factor of two larger than our specification. For SiPM W12 (KETEK), we also overcorrected the bias voltage by $+4.1 \rm mV/^\circ \rm C$ by using $dV/dT=21.3 \rm ~mV/^\circ C$ instead of $17.2  \rm mV/^\circ \rm C$. Thus, for the LCT4 and the W12 SiPMs the stabilization studies need to be repeated with the correct value of $dV/dT$. For The CPTA SiPM, the analysis is still in progress. Dedicated studies of afterpulsing show that gain stabilization is not affected by afterpulsing. We also observe that afterpulsing rises quadratically with the applied overvoltage but is quasi independent of temperature.  

Learning from our experience in August, we carefully planned for a new 10 day stabilization study in the climate chamber at CERN for the middle of  February 2016. We improved our setup to test four SiPMs simultaneously with one bias voltage adjustment using four similar voltage-operational preamplifiers. We transferred light from the LED via individual fibers to each SiPM. For each stabilization run, we took measurements at 20 or more temperatures points between $\rm 1^\circ C-50^\circ C$ with four selected similar SiPMs.  We performed gain stabilization with 30 SiPMS.  From our online analysis we know that all SiPMs performed well  meeting our requirements except for two CPTA sensors. The results will be finalized and published soon. We now have sufficient information to implement a temperature-dependent bias voltage adjustment into the power distribution  system of an analog hadron calorimeter.

\section{Acknowledgment}
This work was conducted in the framework of of the European network AIDA2020.
It has been supported by the Norwegian Research Council and by the Ministry of Education, Youth and Sports of the Czech Republic under the project LG14033. We would like to thank Lucie Linssen, Chris Joram and Wolfgang Klempt for using some of their laboratory and
electronic equipment. We also would like to thank the team of the climate chamber at CERN for support.

\smallskip

\end{document}

%% file: econfmacros.tex
\def\beq{\begin{equation}}
\def\eeq#1{\label{#1}\end{equation}}
\def\eeqn{\end{equation}}

\def\beqa{\begin{eqnarray}}
\def\eeqa#1{\label{#1}\end{eqnarray}}
\def\eeqan{\end{eqnarray}}

\let\bar=\overbar

\def\Dslash{\not{\hbox{\kern-4pt $D$}}}
\def\dslash{\not{\hbox{\kern-2pt $\del$}}}

\def\msb{{\bar{\ssstyle M \kern -1pt S}}}

